\documentclass[12pt]{article}%
\usepackage{amsmath}%
\usepackage{amsfonts}%
\usepackage{amssymb}%
\usepackage{graphicx}
\usepackage{color}
\usepackage{hyperref}
\usepackage[makeroom]{cancel}
\usepackage{authblk}

\usepackage{tikz}
\usetikzlibrary{intersections}
\usetikzlibrary{calc}

\newcommand\cdiag[1]{\vcenter{\hbox{\begin{tikzpicture} 
\coordinate (c) at (0,0);  
\coordinate (cr) at (0.75,0); \coordinate (cu) at (0,0.75); 
\coordinate (cru) at (0.75,0.75); 
\coordinate (cl) at (-0.75,0); \coordinate (cd) at (0,-0.75); 
\coordinate (cld) at (-0.75,-0.75);
 \coordinate (crd) at (0.75,-0.75); \coordinate (clu) at (-0.75,0.75); 
#1 
\end{tikzpicture}}}}

\newcommand{\cK}{\mathcal{K}}
\newcommand{\cR}{\mathcal{R}}

\begin{document}
\title{The dual of non-extremal area: \\
differential entropy in higher dimensions}
\author[1,2]{Vijay Balasubramanian}
\author[2,1]{Charles Rabideau}
\vspace{1cm}
\affil[1] {\small{David Rittenhouse Laboratory, University of Pennsylvania, Philadelphia PA, 19104, U.S.A.}}
\affil[2]{\small{Theoretische Natuurkunde, Vrije Universiteit Brussel (VUB), and
International Solvay Institutes, Pleinlaan 2, B-1050 Brussels, Belgium.}}

\maketitle

\begin{abstract}
The Ryu-Takayanagi formula relates entanglement entropy in a field theory to the area of extremal surfaces anchored to the boundary of a dual AdS space.  It is interesting to ask if there is also an information theoretic interpretation of the areas of non-extremal surfaces that are not necessarily boundary-anchored.  In general, the physics outside such surfaces is associated to observers restricted to a time-strip in the dual boundary field theory.  When the latter is two-dimensional, it is known that the {\it differential entropy} associated to the strip computes the length of the dual bulk curve, and has an interpretation in terms of the information cost in Bell pairs of restoring correlations inaccessible to observers in the strip.  A general realization of this formalism in higher dimensions is unknown.  We first prove a no-go theorem eliminating candidate expressions  for higher dimensional differential entropy based on entropic c-theorems.  Then we propose a new formula in terms of an integral of shape derivatives of the entanglement entropy of ball shaped regions.   Our proposal stems from the physical requirement that differential entropy must be locally finite and conformally invariant.  Demanding cancelation of the well-known UV divergences of entanglement entropy in field theory guides us to our conjecture, which we test for surfaces in $AdS_4$.  Our results suggest a candidate c-function for field theories in arbitrary dimensions.
\end{abstract}

\pagebreak

\tableofcontents

\section{Introduction}

One of the deepest insights of the AdS/CFT correspondence is that scale in the field theory is related to radial position in the $AdS$ dual. There have been many efforts to quantify this relationship by isolating the part of the field theory that describes the interior of a radially bounded region in $AdS$ space, e.g., via analysis of probes \cite{hep-th/9808016,hep-th/9808017,1203.1044,cond-mat/0610375}
, the holographic renormalization group \cite{hep-th/0209067,hep-th/9912012,hep-th/9903190,1010.1264,1010.4036}
, and, most recently, the $T\bar{T}$ deformation \cite{1611.03470}.    A complementary question is to isolate the part of the field theory that describes the {\it exterior} of a radially bounded region in $AdS$ space.  The exterior of a convex bounded region ${\cal R}$ is naturally associated to a time-strip in the dual field theory.   This strip is defined as  the union of the causal developments of the boundary balls that anchor bulk minimal surfaces tangent to ${\cal R}$.

In the $AdS_3$/$CFT_2$ correspondence it is known that the {\it differential entropy} of a time-strip is dual to the length of a bulk curve bounding the infrared region ${\cal R}$ \cite{1305.0856,1310.4204,1408.4770,1409.4473}.  Differential entropy is qualitatively understood as a measure of the uncertainty introduced by restricting the boundary observer to making finite-time measurements.  More precisely, it computes the information cost of restoring correlations that are inaccessible to independent observers making measurements within the time-strip \cite{1410.1540}.  If the bulk curve of interest  lies within an entanglement shadow \cite{1312.3699,1406.5859,1412.5175,1704.03448,1507.00354}, its length may be related to differential entropy computed from generalizations of entanglement  \cite{1406.5859,1609.03991}.  The relation between differential entropy and the length of curves has led to numerous new developments in the AdS/CFT correspondence \cite{1505.05515,1512.01548,1604.03110,1606.03307} which have clarified how $AdS$ geometry and curvature are encoded in quantum entanglement of the dual field theory.

The formalism of differential entropy was generalized to higher dimensions in certain highly symmetric situations \cite{1403.3416,1406.4889}.
  Our aim in this paper to understand the general case without assuming any special symmetries.  Thus we seek an information theoretic quantity in $d+1$-dimensional field theory that is dual to the area of a general convex bounded region ${\cal R}$ in $AdS_{d+2}$, for any $d$.   Following the previous experience in $d=1$, we expect to find a formula in field theory that integrates an expression that measures local area elements of the boundary of ${\cal R}$.   These area elements are finite, so we expect the field theory formula to be built from locally finite quantitites.  In addition, it must be conformally invariant since boundary conformal transformations are simply diffeomorphisms in the bulk and cannot change the area of surfaces.    With these points in mind we consider a family of stationary field theory observers who can make measurements over finite durations before and after a time $t=0$.  Collectively these observers define a time-strip.   The causal diamonds of the worldlines of these observers intersect the $t=0$ surface in ball-shaped regions $B(\sigma)$ where $\sigma$ labels positions in the field theory and the radius of the ball is equal to the length of time the observer has to make measurements. We propose that the differential entropy is given by the integral of certain shape derivatives of the entanglement entropy associated to the family of balls, $B(\sigma)$.
For the simplest extension applicable to $AdS_4$, our proposal is: 
\begin{align}
S_{diff}\Big[\{B(\sigma) \}\Big] &= \int d^2\sigma \mu(\sigma)  \, \delta^{(2)}S
\Big[ \lceil \rho_{T^\leftarrow}\rceil , \lceil\rho_{T^\downarrow} \rceil \Big]\,, \\
\delta^{(2)}S
\Big[ \lceil \rho_{T_1}\rceil , \lceil\rho_{T_2} \rceil \Big]  
&= \int d\theta_1 d\theta_2 \,
\frac{\partial^2 S}{\partial \delta r(\theta_1) \partial \delta r(\theta_2)} 
\lceil \rho_{T_1}\rceil (\theta_1) \lceil\rho_{T_2} \rceil(\theta_2)
\,,
\end{align}
where $\lceil \rho_{T^{(\leftarrow,\downarrow)}}\rceil$ are specific deformations of the shape of the ball determined by the family of balls in question (precise definitions and caveats are given in the main text).  We also consider higher bulk dimensions, which require us to work with higher shape derivatives of the entanglement entropy.   In $d=1$ our expressions reproduce the known formula for differential entropy from \cite{1310.4204}.

We start in Section \ref{sec:maths}, by introducing a mathematical tool -- double-fibrations -- that are essential in the study of integral geometry.
We then apply this language to the problem at hand by constructing the double-fibration relating an arbitrary bulk and its associated boundary kinematic space of ball-shaped regions. This kinematic space is a natural generalisation of the space of intervals that was used to define time strips and the differential entropy in $AdS_3/CFT_2$ and is equivalent to the set of ball-shaped regions $B(\sigma)$ introduced above.  This language allows us to prove a no-go theorem ruling out some of the simpler proposals for differential entropy that have appeared in the literature.  With this result in hand, in Section \ref{sec:diff_ent}, we develop our proposal for higher dimensional differential entropy of a family of balls $B(\sigma)$, which we conjecture is dual to the area of an associated bulk surface bounding an interior region ${\cal R}$.  In Section \ref{sec:verify}, we verify that this conjecture holds in the vacuum for $d=2$, i.e. in $AdS_4$/$CFT_3$.  More general tests will require new computations of shape derivatives of entanglement entropy to higher orders or in excited states. Finally, in Section \ref{sec:discuss} we discuss a possible interpretation of our higher dimensional generalisation of differential entropy in terms of the cost in Bell pairs of a constrained state merging protocol which reconstructs the entanglement of a global state from the restriction of the state to each ball and make a proposal for extracting c-functions in any dimension from the shape derivatives appearing in our higher dimensional generalisation of differential entropy.

\section{Mathematical arena}\label{sec:maths}
In the body of this work, we will consider states of a $CFT_{D=d+1}$ with time-reflection symmetry across a $d$-dimensional spatial slice on which the Ryu-Takayanagi (RT) formula for holographic entanglement entropy applies \cite{hep-th/0603001}.\footnote{In this work we will use $D$ to denote boundary spacetime dimensions and $d$ to denote the spatial dimensions, since we will be working on a spatial slice. Much of the following can be covariantly generalised using the Hubeny-Rangamani-Takayanagi extension of the RT formula \cite{0705.0016}, but since this makes the notation much heavier we will deal with the general case in Appendix~\ref{sec:lorentzian}.}   Since we are working in a time-refelection symmetric setting the extremal RT surfaces in $AdS_{D+1}$ that compute entanglement entropy will be restricted to a spatial slice of the geometry.  This slice is  is a  $(d+1)$-dimensional asymptotically hyperbolic Riemannian manifold $M$. We will refer to any co-dimension one submanifold of $M$ as a ``surface''.

On $M$ the volume $(d+1)$-form, $vol$ computes the volume of co-dimension 0 subregions of $M$:
\begin{align}
Vol( X \subset M ) = \int_X vol \,.
\end{align}
Meanwhile, the area of a co-dimension one surface $N$ embedded in $M$ via $x^a(\sigma^i)$ can be written
\begin{align}
Area(N) = \int_N d^d \sigma \sqrt{det_{ij} g_{ab} \partial_i x^a(\sigma) \partial_j x^b(\sigma)} \,.
\end{align}
This formula, which we seek to reconstruct from boundary quantities, depends on derivatives of the embedding and so it is not simply the integral of a fixed $d$-form which can be chosen independently of $N$. In other words, the quantity which we must integrate to obtain the area depends on the tangents to the surface at each point on the surface in addition to metric at that point.
This implicit dependence on the tangents to the surface can be written in terms of the normal vector $n$ contracted into the volume form as
\begin{align}
\label{eqn:area_norm}
Area(N) = \int_N \iota_n vol \,,
\end{align}
where $\iota_V \omega$ denotes the contraction of the vector $V$ into the form $\omega$.
By appending to our surface $N$ the information about its normal vectors, we can make this dependence explicit.

The addition of the normal vector data to each  point on the surface $N$ is conveniently described in terms of a section of the fibre bundle of unit vectors:\footnote{A fibre bundle $E$ is a manifold which locally has the product structure $M \times F$, so that we have associated a space $F$ called a fibre to each point of the base manifold $M$. This means that points $e \in E$ can be thought of as pairs $e = (p\in M,f\in F)$.
Formally, a bundle is defined by a triple $(E,M,\pi)$, where $\pi:E \rightarrow M$ is the projection of the bundle $E$ onto the base $M$ which consists of forgetting the fibre. The fibre above a point $p \in M$ is $\pi^{-1}(p)$. A section of a bundle is a map $s: M \rightarrow E$, such that $\pi(s(p))=p$ for all $p \in M$. This means that it assigns a particular point in the fibre to each point on the base. In this work we will also call a submanifold $\tilde N$ of $E$ a section if it picks out a unique point in each fibre above the base of the submanifold, $N \equiv \pi(\tilde N)$. In other words, $\tilde N \subset E$ will be called a section if $(\pi |_{\tilde N})^{-1}$ is a section. }
\begin{align}
\mathbb{S} M = \{ (x,V) \in TM | g_{ab} V^a V^b =1 \} \,, 
\end{align}
where $TM$ is the tangent bundle of $M$.
The fibres of this bundle can be thought of as spheres living at each point of $M$. 
There is a natural lift of the surface $N$ into a section of $\mathbb{S} M $ which we call $\tilde{N}$, by adjoining the normal vector to each point on the surface.

This bundle allows us to define an ``area'' form\footnote{This area form defines a signed area based on the assignment of the sign of the normal vector when lifting from $N$ to $\tilde{N}$. An unsigned area could be defined by  integrating the absolute value of this quantity. In \cite{1505.05515} the considering a signed area was important for reconstructing complex surfaces, but in this work we will only work locally and so we will not worry about this sign.}
\begin{align}
area = \iota_V vol
\end{align}
as a $d$-form on $\mathbb{S} M$ 
by contracting the vector from the fibre of $\mathbb{S}M$ into the volume form on $M$.
When integrated over $\tilde N$, this area form reproduces \eqref{eqn:area_norm}.

Let us unpack this definition for the simple case of empty $AdS_{D+1}$ with a spatial slice $M = \mathbb{H}_{d+1}$. 
Using Poincar\'e coordinates, 
\begin{align} \label{eqn:Poincare_coords}
ds^2 = \frac{dz^2 + dx_i^2}{z^2} \,,
\end{align}
the embedding of a surface $N$ can be described locally by specifying it as the level set $z = z(x_i)$. The area of this surface is given by computing the induced metric
\begin{align} \label{eqn:AdS_area_1}
ds^2_\mathrm{ind} = 
\frac{\delta_{ij} + (\partial_i z)  (\partial_j z) }{z^2} dx_i dx_j \,, \\
Area(N) = \int d^d x \, z^{-d} \sqrt{1 + (\partial z)^2} \,,
\end{align}
where $\partial_i z = \frac{\partial z (x_i)}{\partial x_i}$.

The normal vector to a level set is in the direction of the gradient obtained from the exterior derivative $d \big( z - z(x_i) \big)$. However, this is a 1-form, so to make a vector we must raise the index with the metric and then normalise it to obtain the unit normal:
\begin{align}
d \big( z - z(x_i) \big) &= dz - \partial_i z(x_i) \, d x_i \,, \\
n^a &=  \frac{z ( 1, -\partial_i z )}{\sqrt{1 + (\partial z )^2}} \,,
\end{align}
where we have used notation where a vector $a \partial_z + b^i \partial_{x^i}$ is denoted $(a,b^i)$. 

A unit vector at a point $(z,x_i)$ on $M$ can be parametrized (up to an overall sign) by $d$ numbers $\dot z_i$,
\begin{align}
V^a &=  \frac{z ( 1, -\dot z_i )}{\sqrt{1 + (\dot z )^2}} \,,
\label{Ecoords1}
\end{align}
which gives a coordinate system, $(z, x_i, \dot z_i)$, on the bundle of unit vectors 
$\mathbb{S} M$. 
The lift of the surface $N$ into a section of this bundle, $\tilde N$, is given by setting $\dot z_i = \partial_i z$.

In these coordinates the area form is
\begin{align}
area &\equiv \iota_V vol = \frac{dx_1 \wedge \ldots \wedge dx_d + \sum_{i=1}^d dx_1 \wedge \ldots \big\{
\cancel{dx_i} \rightarrow ( \dot z_i dz ) \big\}
 \ldots \wedge dx_d}{z^{d} \sqrt{1 + \dot z^2}} \,. \label{eqn:area_form}
\end{align}
where the notation in the sum means that we should replace $dx_i$ by $\dot{z}_i dz$ in the the $i^\mathrm{th}$ term.
This is a well defined $d$-form on 
$\mathbb{S} M$. 
Integrating this form on $\tilde N$ corresponds to setting $z=z(x_i)$ and $\dot z_i = \partial_i z$, which reproduces \eqref{eqn:AdS_area_1}.

In this language, {\it  identifying the boundary dual of the area of a surface in the bulk is equivalent to the question of finding the dual of the area form.} This formulation makes explicit that the integrand in the area formula is a local quantity and that reproducing this form up to a total derivative is sufficient to capture the area of a closed surface. In the following subsection we will reformulate this question in terms of information theoretic boundary CFT quantities.

\subsection{Kinematic space}
Entanglement entropy in the CFT dual to asymptotically $AdS$ space is a function on the set of subregions of the field theory.  In two dimensional CFTs this is simply the space of intervals, identified by the authors of \cite{1505.05515} as the ``kinematic space'' of the theory.  We will consider the generalised kinematic space $\cK$ introduced in \cite{1509.00113,1604.03110} and further studied in \cite{1606.03307,1812.02176}. This can be thought of as the manifold of ball shaped regions in a spatial slice of our $CFT_{D=d+1}$. When this spatial slice is $\mathbb{R}^d$ we can parametrize the $d+1$-dimensional space of balls $\cK$ by the center and radius of each ball, $(R,x_0^i)$.  This is of course a subset of the set of all subregions on which entanglement entropy is defined, but it will suffice for our purposes.

Another reason to be interested in ball shaped regions is that they correspond to the regions accessible to observers who only have a limited time to make measurements. Consider an observer at a fixed position making measurements for a time interval $t\in [-T,T]$. This observer has access to a causal diamond which is the causal development of a ball shaped region centered at the position of the observer with radius $T$. An observer who is allowed to move along a timelike trajectory for a fixed proper time has access to a causal diamond whose tips are not at the same position. However, the tips can be brought back to the same position by a boost and in that boost frame the diamond will again be the causal development of a ball shaped region. Therefore the region accessible to a general observer with a finite lifetime is a causal diamond associated to a \textit{boosted} ball shaped region. This is discussed in more detail in the works \cite{1604.03110,1606.03307}. In the body of this work, we restrict ourselves to ball shaped regions on a fixed time slice for notational simplicity. The formalism of this section is extended to the full covariant kinematic space of general causal diamonds corresponding to boosted ball shaped regions in Appendix \ref{sec:lorentzian}.

The RT formula associates a bulk extremal surface anchored to each of the balls in the kinematic space $\cK$. This extremal surface has co-dimension one in the bulk spatial slice $M$ and is topologically a $d$-dimensional disk $D_d$.
Kinematic space can be augmented by adding a fibre above each point consisting of this extremal RT surface.  In other words, we can introduce a bundle $E$ which is locally $\cK \times D_d$. The bundle $E$ is equivalently the space of all the points on all RT surfaces,   where the point in $\cK$ keeps track of which RT surface we are referring to, while the fibre $D_d$ encodes all the points on the surface. In other words, points in the bundle are  pairs $(k\in \cK,x \in M)$ where the $x$ is a point in the extremal surface anchored at $k$.    We will only use local properties of this bundle in this work.

We can specify coordinates on $\cK$ by stating the radius and center of the ball in question, $(R,x_0^i)$.  The extremal surface anchored on this ball  is topologically a disk, and hence we can specify a point on it by specifying angles $\Omega_j$ in an upper hemisphere (which projects down to a disk). Then $(R,x_0^i,\Omega_j)$ provide coordinates on $E$. To summarise, we have the bundle structure
\begin{align}
\pi_\cK : E \rightarrow \cK \,, \qquad \pi_\cK^{-1}(p\in \cK) = D_d \,,
\end{align}
where the projection $\pi_\cK$ consists of forgetting where on the RT surface the point lies and keeping track only of the whole surface.  This bundle also has another projection, $\pi_M$, which consists of forgetting which RT surface a point in $E$ lies on and keeping track of it as just a point in $M$. This gives $E$ the structure of a double-fibration. 

Given a point on an extremal surface, $p\in E$, we can associate to it the normal vector to the extremal surface at that point. This provides a natural embedding of $E$ into the bundle of unit vectors on $M$, which we called $\mathbb{S}M$. In other words, the bundle $E$ can equivalently be thought of as a submanifold of $\mathbb{S}M$. In fact, for the case of empty $AdS$ they are equal because every point has a minimal surface passing through it with a unit normal pointing in any given direction, as will be demonstrated in Section \ref{sec:embed}.
However in general, due to the presence of entanglement shadows \cite{1312.3699,1406.5859,1703.07780}, this will not be the case. 

Understanding the manifestation of entanglement shadows in this language will be an important future problem, but again we will only use the local properties of the bundle $E$ in this work and so we will not worry about this yet. The projection that $E$ inherits from the embedding $E \rightarrow \mathbb{S}M$ is exactly the projection $\pi_M$ and so we can also try to think of $E$ as a bundle with base $M$ and fibres which are topologically a unit sphere:
\begin{align}
 \pi_M: E \rightarrow M \,, \qquad \pi_M^{-1}(p\in \cK) \subset S_{d} \,.
 \end{align}
In general the fibres will only be subsets of $S_d$ since not every point of $M$ necessarily has boundary anchored extremal surfaces of minimal area passing through it with normals in every direction.
This means that $E$ is not really a bundle with respect to the base $M$, since the fibres will include different subsets of the sphere, although it remains a bundle with respect to the base $\cK$. The general statement is simply that $E$ can be embedded locally into $\mathbb{S}M$.

A co-dimension 1 submanifold $J$ of $\cK$, which we refer to as a family of balls, defines a family of bulk RT surfaces and the envelope of this family defines a bulk co-dimension 1 submanifold or surface $N \subset M$. As long as the family of balls $J$ is sufficiently ``nice''\footnote{The condition for a family of balls $J$ to be considered ``nice'' is given in \eqref{eqn:norm_R_condition} once we have developed the required machinery.}, each extremal surface will be tangent to the bulk surface $N$ at a point and their normal vectors will be the same, thus there is a natural way to lift $J$ to a section $\tilde J \cong \tilde N$ of $E$.  By thinking of a ball shaped region as the region accessible to an observer in the boundary field theory at a fixed position making measurements for a finite time, a family of balls can also be thought of as the time strip accessible to a family of observers. A family of balls parametrized  by  $R(x_0^i)$ corresponds to a time strip where observers at $x_0^i$ can make measurements in an interval $t\in  [-R(x_0^i),R(x_0^i)]$. A general time strip $t\in [t_-(x_0),t_+(x_0)]$ with $t_- \neq t_+$ corresponds to measurements accessible to observers living in causal diamonds which are the causal developments of the boosted balls considered in Appendix~\ref{sec:lorentzian}.

Using the language developed in this section, the boundary dual of the area of a bulk surface is given by a prescription starting in $\cK$ (the space of balls in the CFT) to reconstruct the area form on the bundle $E$ (the space of points on bulk extremal surfaces) thought of as a subset of $\mathbb{S}M$ (the space of unit vectors on a bulk spatial slice) without assuming knowledge of the metric on the bulk spatial slice $M$.

\subsection{No-go theorem}
In order to connect the formalism developed above to previous work, we will rewrite the differential entropy formula for reconstructing the length of curves in $AdS_3$ \cite{1310.4204,1505.05515}. We will see that there is a significant simplification in this low-dimensional case.
Writing things in our formalism will allow us to rule out a certain class of possible extensions of the differential entropy to higher dimensions, which includes the proposal given for example in \cite{1805.08891} based on holographic c-functions. 

In three bulk dimensions, so that in our notation $d=1$, \cite{1505.05515} showed that  the area can be computed from a form $c$ defined on $\cK$, which we will call the Crofton potential.\footnote{The Crofton form, which is the quantity more often discussed in integral geometry, is the differential of the Crofton potential \cite{1505.05515}. See also Appendices \ref{app:croft} and \ref{app:int_geo} for a discussion of how to apply ideas from integral geometry in our framework.} This potential can be obtained by taking a derivative of the entanglement entropy of a ball shaped region $S(R,x_0)$
\begin{align}
c &=  \partial_R S(R,x_0) \, dx_0 \, , \\
Area &= \int_{\tilde J} area  = \int_J c \,.
\end{align}
In the second line the Crofton potential is integrated over a family of balls $J$ (i.e. a subset of kinematic space) to get the area of the envelope of the extremal surfaces anchored to the balls in $J$. This is a surprising relation in light of the structure we have constructed so far: how is it that this form $c$ lives directly on $\cK$ rather than on $\mathbb{S}M$ where the area form is supposed to live? This is made possible by the fact that in three dimensions the area form  does not depend on the fibres of $E$ when thinking about it as a bundle over  $\cK$. 

To show this, we start by noting that a spatial slice of $AdS_3$ in Poincar\'e coordinates has metric and area form
\begin{align}
ds^2 &= \frac{dz^2 + dx^2 }{z^2} \,. \\
area &= \frac{dx + \dot z dz}{ z \sqrt{1+\dot z^2}} \,. \label{eqn:AdS_area_2}
\end{align}
The associated Crofton potential is defined on the space of intervals on the boundary of $AdS_3$, namely the kinematic space parametrized by the center and width of the interval, $(R,x_0)$.    For a CFT in the vacuum, the entanglement entropy $S$ and associated Crofton form are then
\begin{align}
S(R,x_0) &= \log{R}{\epsilon} \,, \\
c &= \frac{dx_0}{R} \,.
\end{align}
We now want to construct the bundle $E$ of points on extremal surfaces. To give coordinates on $E$ we must augment the coordinates on kinematic space, which specify which extremal surface we are working with, by specifying a particular point on the surface.   Boundary-anchored extremal surfaces in $AdS_3$ are  defined by the equation $z^2+(x-x_0)^2 = R^2$. Thus, a good set  of coordinates on $E$ is $(R,x_0,s)$ with $s = \frac{x_0-x}{z}$.

To check the Crofton formula for area, we can now simply change coordinates on the bundle $E$ from $(z,x,\dot z)$ as given in (\ref{Ecoords1}) to the new coordinates $(R,x_0,s)$. We can relate these coordinate systems  through a parametric description of the extremal surfaces as
\begin{align}
z(x) &= \sqrt{R^2 - (x-x_0)^2}  \implies z'(x) = \frac{x_0-x}{z} \,.
\end{align}
The normal vector attached to the point on the extremal surface parametrized by $(R, x_0, s)$ is $(z,x,\dot z)$ with $\dot z = z'$.
Explicitly, then
\begin{align}
\dot z =  s  ~~~~;~~~~
z = \frac{R}{\sqrt{1+s^2}}  ~~~~;~~~~
x = x_0 - R \frac{s}{\sqrt{1+s^2}} \,.
\end{align}
and the inverse transform is
\begin{align}
R = z \sqrt{1+\dot z^2}  ~~~~;~~~~
x_0 = x + z \dot z \,.
\end{align}
Applying this to the area form in \eqref{eqn:AdS_area_2}
\begin{align}
area &= \frac{d\big( x_0 - R \frac{s}{\sqrt{1+s^2}}\big) + s d\big( \frac{R}{\sqrt{1+s^2}} \big)}{R} \,, \\
&= \frac{dx_0}{R}   - d \big(\sinh^{-1} s \big) = c - d \big(\sinh^{-1} s \big) \,.
\end{align}
Therefore we see that the Crofton potential reproduces the area up to boundary terms.\footnote{In fact, in this case the boundary term can even be understood as computing the contribution to the area of the envelope coming from following the last extremal surface in the family down to the boundary.}

The computation above worked because the area form did not depend on $s$, i.e. it was not a function of location on the extremal surface.  However, in higher dimensions, an explicit computation of the area form for empty $AdS$ shows that this property no longer holds: the area form depends on the position on the extremal surface.   
To see this, recall that in any dimension the area form for empty $AdS$ is \eqref{eqn:area_form}.    By following the same procedure as above, the area form can be rewritten in coordinates on $E$: $(R,x_0^i,s^i)$, where $s^i = \frac{x^i_0 - x^i}{z}$.  
An explicit transformation to these coordinates yields a complicated expression (not shown) which depends on the $s^i$.\footnote{That the dependence on the $s^i$ is not a boundary term can be verified by checking that $d (area)$ has a non-trivial dependence on $s^i$.}   
Therefore simply integrating a form in kinematic space cannot reproduce the area of all surfaces since the kinematic space only encodes the extremal surfaces and not the points on them.  
We can state this as a {\bf no-go theorem}:  For $\mathbb{H}_{d+1}$ with $d>1$, there exists no $d$-form $c$ on $\cK$ (the space of boundary balls) such that the integration of $c$ over an arbitrary family of balls reproduces the area of the envelope of the extremal surfaces in $\mathbb{H}_{d+1}$ anchored to those balls.

In particular, this rules out the hope, based on area theorems \cite{1805.08891} or analogies with the differential entropy in 2d CFTs, that quantities appearing in the proofs of entropic $c$-theorems \cite{cond-mat/0610375,1506.06195,1704.01870} could provide us with a higher dimensional Crofton potential which computes areas in AdS.

 To explore this in more detail, start with the case of $d=1$ ($CFT_2$/$AdS_3$), and a geometry of the form
\begin{align}
ds^2 = \frac{  dz^2 + e^{2 A(z)} dx^2}{z^2} \,,\\
 \end{align}
dual to the RG flow of a field theory. In such a geometry, the monotonicity of $A(z)$ appearing in the metric follows from the null energy condition and so $A(z)$ can be thought of as a c-function \cite{hep-th/9904017,1006.1263,1011.5819}. The area of the bulk surface at fixed $z=z_0$ measures this function and so provides a geometric c-function for this RG flow. From the boundary, a c-function can be constructed from the entanglement entropy of an interval by taking the appropriate derivatives to extract the universal term \cite{cond-mat/0610375}:
\begin{align}
c(R) = R \frac{\partial S(R)}{\partial R} \,.
\end{align}
The differential entropy relates these two approaches: in a translation invariant geometry, a surface at fixed $z=z_0$ is the envelope associated to the set of all intervals of a given size $R=R_0$. Thus
\begin{align}
\frac{1}{4 G_N} Area(R_0) = \int dx_0 \left. \frac{\partial S(R,x_0)}{\partial R} \right|_{R=R_0} = \frac{2\pi L}{R_0} c(R_0) \,, 
\end{align}
where the field theory lives on a circle of radius $L$.

In higher dimensions the area of a surface at $z=z_0$ still provides a holographic RG monotone \cite{1805.08891}. On the boundary, RG monotones can be constructed from the entanglement entropy of ball shaped regions. In particular, in $CFT_3$/$AdS_4$ \cite{1506.06195},
\begin{align}
C(R) = \frac{1}{2\pi} \Big[ R \frac{\partial S}{\partial R} - S \Big]\,,
\end{align}
is an RG monotone.\footnote{Similar expressions constructed from derivatives of the entanglement entropy were proposed by \cite{1202.2070} to be candidates for RG monotones in arbitrary dimensions.}
This leads to the hope that these two approaches could be related by a differential entropy formula of the form:
\begin{align}
Area\big( R(x_0^i) \big) \stackrel{?}{=} \int d^d x_0^i  \, C\big(R(x_0^i) \big) \,,
\end{align}
where the area of a bulk surface would be dual to a particular combination of derivatives of the entanglement entropy of balls, integrated over the family of balls tangent to the bulk surface. 

However, such a formula is the integral of a form defined on $\cK$ and so is ruled out by our no-go theorem. Of course, certain classes of surfaces might be reconstructed in this way. In particular, if we restrict attention to surfaces which lie on particular sections of $E$, such as the surfaces at constant $z=z_0$ in a translation invariant geometry, the area form naturally restricts to a form on $\cK$.   This is also why, in the translation invariant case, the authors of \cite{1403.3416} were able to generalize differential entropy directly to higher dimensions. Nonetheless, the argument leading to our no-go theorem demonstrates why this approach cannot work to give the dual description of the areas of all the surfaces in a given geometry.    Faced with this no-go theorem, we will propose another approach to generalising differential entropy in Section \ref{sec:diff_ent}.   This new approach is easiest to understand in an embedding space approach to $AdS$ geometry which we will set up in the remainder of this section.

\subsection{Embedding space}
\label{sec:embed}
The symmetries of a CFT can be linearised by introducing an embedding space, which provides a natural arena for understanding the various manifestations of the bundle $E$ described in the previous section. Many of the computations in this work will be done in embedding space, so we will here establish our conventions and rephrase the mathematical framework we have developed so far in this language.

For a $d$-dimensional spatial slice of a CFT, the embedding space is a $d+2$ dimensional vector space where the $SO(d+1,1)$ subgroup of the conformal group preserving this spatial slice acts in the defining representation. Points in the CFT are encoded in null vectors $Z^2=0$ modulo positive rescalings $Z \cong \lambda Z$ with $\lambda>0$. 
We will write the metric with signature $(-,+,+\ldots)$. Considering a CFT whose spatial slices are $\mathbb{R}^{d}$ corresponds to parametrizing $Z$ by\footnote{We use indices $i,j,k =1\ldots d$ for boundary spatial coordinates, $a,b,c =1 \dots d+1$ for bulk coordinates and $A,B,C=1\ldots d+2$ for embedding space coordinates.}
\begin{align}
Z = \Big( \frac{1 +x^2}{2}, x^i, \frac{1-x^2}{2}  \Big),
\label{eqn:embed_planar}
\end{align}
where $i=1\ldots d$.

The space of balls, $\cK$, was described in this language in \cite{1606.03307}.  A ball in the CFT can be parametrized in terms of a unit spacelike vector $B^2=1$: the points within the ball are those with $Z \cdot B > 0$. Direct calculation confirms that a ball with center $x_0^i$ and radius $R$ is corresponds to
\begin{align}
B = \Big( \frac{1- (R^2 - x_0^2)}{2 R}, \frac{x_0^i}{R}, \frac{1+ (R^2-x_0^2) }{2 R} \Big).
\end{align}

We can also use the embedding space formalism to describe the geometry of empty $AdS$ space, although it should be emphasized that embedding space encodes the symmetries of a $CFT$ independent of any dual $AdS$. Points in empty $AdS$ correspond to unit timelike vectors $X^2=-1$. Poincare coordinates correspond to 
\begin{align}
X = \Big( \frac{1 +(z^2 + x^2)}{2z}, \frac{x^i}{z}, \frac{1-(z^2+ x^2)}{2z}  \Big).
\end{align}
The minimal surface anchored on the boundary of a ball $B$ is a hemisphere encoded as the points such that 
\begin{align}
 X\cdot B=0\,.
 \end{align}
In other words, the intersection relation between points in $AdS$ and extremal surfaces anchored on the boundary of a ball is replaced by an orthogonality relation.

A point on a given extremal surface, that is a point in the bundle $e \in E$ associated to empty AdS, is described by a pair $e=(X,B)$ such that $X \cdot B=0$, $X^2 = -1$ and $B^2 = 1$. In terms of the double-fibration structure, $\pi_M(e) = X$ and $\pi_\cK(e) = B$.

Now we will see that a vector $B$, encoding a ball whose extremal surface is tangent to a bulk surface $X(\sigma)$, is in fact simply the normal to that surface in embedding space. In this language the embedding of $E$ into the unit vector bundle $\mathbb{S}M$ is trivial.

Let $X(\sigma)$ be a surface in $AdS_{D+1}$. Let $Y(\sigma)$ be an extremal surface anchored on a ball tangent to the surface at a given point $X(\sigma_0)$, so that 
\begin{align}
Y(\sigma_0) &= X(\sigma_0) \,, \\  \partial_{\sigma^i} Y(\sigma_0) &= \partial_{\sigma^i} X(\sigma_0)\,.
\end{align}
Denote by $B$ the ball on which the extremal surface $Y(\sigma)$ is anchored, so that $B\cdot Y(\sigma)=0$ and therefore 
\begin{align}
 \partial_\sigma \big( B \cdot Y(\sigma) \big) = B \cdot \partial_\sigma Y =0 \,.
 \end{align}
In other words, $B$ is also orthogonal to the tangent space of the extremal surface which is anchored to it. Since $Y$ is tangent to $X$ at the point $X(\sigma_0)$, this implies that $B$ is also orthogonal to the tangent space of the bulk surface $X$ at this point. A surface $X$, which has co-dimension 1 in the bulk spatial slice, has a $d$-dimensional tangent space, so orthogonality with $X$ itself in addition to the tangent space is $d+1$ conditions in the $(d+2)$-dimensional embedding space.  These conditions are sufficient to fix $B$ (up to a sign) to be the unit normal vector to the surface.

Now let $B(\sigma)$ be the ball or normal vector associated to each point on $X(\sigma)$. Since $B(\sigma) \cdot X(\sigma) =0$ for all $\sigma$, 
\begin{align}
\partial_{\sigma^i} B \cdot X = - B \cdot \partial_{\sigma^i} X =0 \,.
\end{align}
So, the tangent plane of the family of balls is the same as the tangent plane of the bulk surface -- it is the space orthogonal to both $B$ and $X$. This also gives a third interpretation of the bundle $E$ in terms of the unit vector bundle on $\cK$, where the fibre is given by the vector $X$ and the metric on $\cK$ is that inherited from embedding space or in other words it is the unique conformally invariant metric on $\cK$. 

In this embedding space language, all the different interpretations we have given to our bundle come back to the fact that the pair $(X,B)$ can be interpreted in many ways. $B$ can be either a normal vector to a surface in $AdS$ or it can specify a ball in the dual CFT. $X$ can either specify a point in $AdS$ or it can specify the normal to a family of balls. Working in embedding space allows us to move effortlessly between these different interpretations.

\subsection{Integral geometry and empty $AdS$} \label{sec:int_geo}

The field of integral geometry tells us that sufficiently symmetric spaces can have uniquely defined measures consistent with the symmetry on geometric objects, like geodesics or k-planes. This has limited utility for the general problem of reconstructing geometry from entanglement in higher dimensions, but since empty $AdS_{D+1}$ is a homogeneous space we can use integral geometry to define invariant measures which will compute the area of surfaces in it.

This is easiest to understand in embedding space.  In this presentation, a spatial slice of $AdS_{D+1}$ is the set of unit timelike vectors, $X^2=-1$. A natural way to construct a measure is to start with a metric and then introduce the volume form. The metric invariant under $SO(d+1,1)$ is 
\begin{align}\label{eqn:AdS_metric_embed}
ds^2 = dX \cdot dX \,.
\end{align}
If we expand this in  terms of the parametrization of a Poincar\'e patch in \eqref{eqn:embed_planar}, we recover the usual Poincar\'e metric.
Similarly, we can introduce an invariant metric and measure on $\cK$:
\begin{align}\label{eqn:K_metric_embed}
ds^2_\cK &= dB \cdot dB \,.
\end{align}
The resulting metric on $\cK$ is dS$_{d+1}$ \cite{1505.05515,1509.00113,1604.02687,1606.03307}.

Given a family of balls on the boundary, $B(\sigma)$, we can construct the unique invariant measure by considering the metric induced from \eqref{eqn:K_metric_embed}. This family of balls also specifies a bulk surface $X(\sigma)$, i.e. the envelope of the extremal bulk surfaces anchored to the ball.   Then $X(\sigma)$ also has a unique invariant measure given by the metric induced from \eqref{eqn:AdS_metric_embed}. Since these are the unique measures we can construct on this space that respect the  $SO(d+1,1)$ symmetry, they must be equal up to normalisation (and up to a possible boundary term which we have not tracked in this argument). 
In other words, integrating the measure obtained from (\ref{eqn:AdS_metric_embed}) over $X(\sigma)$ should give the same results as integrating the measure obtained from \eqref{eqn:K_metric_embed} over $B(\sigma)$. This is checked by explicit computation in Appendix \ref{app:croft}.

This argument gives us a formula for reconstructing the area of surfaces in a spatial slice of empty $AdS_{D+1}$ purely from the boundary data.  Simply integrate the invariant measure on $\cK$ over the balls whose extremal surfaces are tangent to the desired bulk surface.   However, this does not give us any indication of how to reconstruct the area of surfaces in a more general asympotically $AdS$ geometry, since it was not constructive and relied entirely on symmetry to tell us that two seemingly unrelated quantities had to be the same since there was only one structure allowed by the symmetry.\footnote{This symmetry based approach has quite general applicability to empty $AdS_{D+1}$. Many geometric questions, such as finding the volume enclosed by a surface or the areas of arbitrary co-dimension surfaces, can be attacked with this approach. Examples of this approach can be found in the literature in integral geometry \cite{helgason,santalo_kac_2004,solanes}, which can be supplemented by embedding space which provides a tool for computing the required invariant measures in $AdS_{D+1}$. See Appendix \ref{app:int_geo} for a discussion of this approach.}
In the following section, we will introduce another approach to this problem which does not rely on this symmetry.

\section{Differential entropy}
\label{sec:diff_ent}
Now we return to the problem of generalising differential entropy. Start by recalling the setup in $d=1$ (i.e., $(1+1)$-dimensional field theories).
Let $S(A)$ denote the area of the extremal surface anchored to a boundary region $A$, and consider the bulk surface which is the envelope of extremal surfaces anchored to the family of intervals $I(\sigma)$. Then differential entropy is the continuum limit of 
\begin{align}
S_{diff} [I(\sigma)] = \sum_{i=1}^N \left[ S(I_i) - S(I_i\cap I_{i-1}) \right],
\end{align}
where $I(\sigma_i) = I_i$ is a discretization of this family of intervals.
We have defined $I_0 = \emptyset$, although in what follows we will not worry about such boundary terms.

Denoting the interval $I_i$ by its endpoints $[u_i,v_i]$, so that $I_i\cap I_{i-1} = [u_i,v_{i-1}]$. The entanglement entropy function can be expanded in a series as 
\begin{align}
S(I_i\cap I_{i-1}) &= S( [ u_i,v_{i-1}]) \\
&= S( [ u_i,v_{i}]) - (v_i - v_{i-1}) \left[ \partial_v S([u,v]) \right]_{u=u_i,v=v_i} + O\left( v_i-v_{i-1} \right)^2, \nonumber
\end{align}
so that in the continuum limit we recover the usual differential entropy formula
\begin{align}
\sum_{i=1}^N \left[ S(I_i) - S(I_i\cap I_{i-1}) \right] = \sum_{i=1}^N \Delta v_i \left[ \partial_v S([u,v]) \right]_{u=u_i,v=v_i} = \int du  \left[ \partial_v S([u,v]) \right]_{v=v(u)} \,. \nonumber
\end{align}

In $d=1$, the space of all connected boundary regions is simply the space of intervals. 
In higher dimensions, the space of all connected regions on the boundary is much more complicated. 
However, the space of intervals can also be thought of as the $d=1$ case of the space of balls $\cK$.
Since this space played an important role in our mathematical description of the problem of reconstructing the area of a bulk surface, we will generalise the space of intervals to this space of balls $\cK$ when constructing a higher dimensional generalisation of the differential entropy.

\begin{figure}[t]
\centering
\includegraphics[width=.7\textwidth]{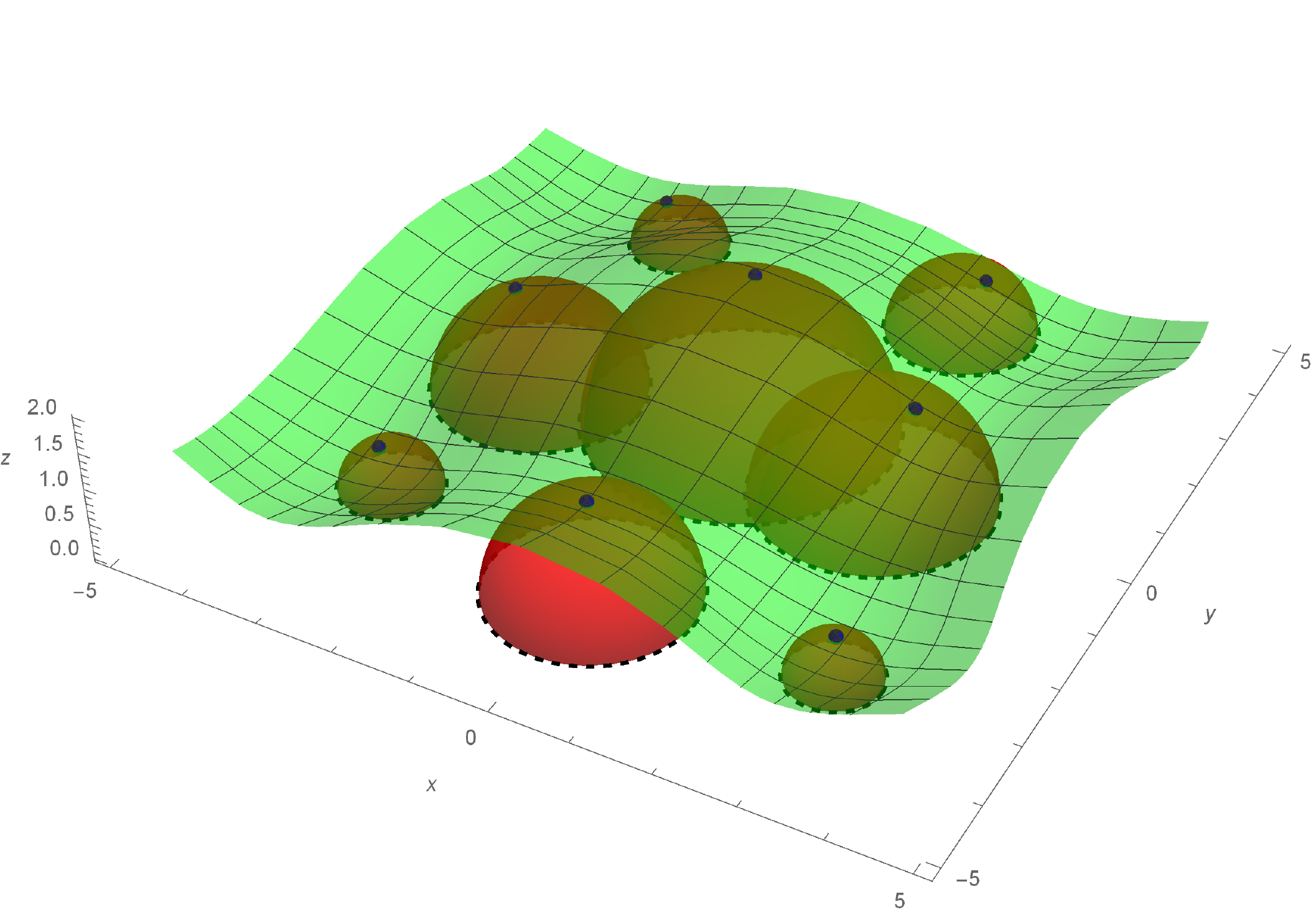}
\caption{The envelope (in green) of the extremal surfaces (in red) anchored to the family of balls $R(x_0,y_0) = 2 - \frac{1}{\sqrt{2}}\left( \sin^2(x_0/2) + \sin^2(y_0/2) \right)$ in a spatial slice of empty $AdS_4$. Poincar\'e coordinates (defined in \eqref{eqn:Poincare_coords}) are used, therefore the boundary of the spatial slice is at $z=0$. The blue points highlight where the extremal surfaces depicted on the plot are tangent to envelope and the dashed lines are the boundaries of the balls to which the extremal surfaces are anchored.}
\label{fig:envelope}
\end{figure}

Therefore, the generalisation of the differential entropy we wish to construct is the boundary dual to the area of a bulk surface defined as the envelope of the extremal surfaces anchored to a family of balls $B(\sigma)$ in $\cK$. An example of such an envelope can be found in Figure~\ref{fig:envelope}.

We will be guided by two criteria that this quantity must obey:
\begin{enumerate}
\item The area of a bulk surface only has divergences where this surface approaches the asymptotic boundary of the bulk. The envelope associated to a family of balls only approaches this asymptotic boundary at the boundary of the family. If the family of balls is a closed surface with no boundary, such that the bulk envelope is a closed surface inside global AdS, then the area is finite. Thus the differential entropy must be finite in the interior of the family of balls, with divergences only allowed near boundaries.
\item A change of conformal frame on the boundary is dual to a change of coordinates near the asymptotic boundary in the bulk. The area of a bulk surface is invariant under a change of coordinates, so the differential entropy must be invariant under conformal transformations of the boundary.
\end{enumerate}

We claim that the appropriate generalisation of differential entropy to $d=2$ (i.e., $(2+1)$-dimensional field theories) is
\begin{align}
S_{diff}[ \{ B_{i,j} \}] = \label{eqn:diff_entropy} 
\sum_{i,j=1}^N \Bigg[ S&(B_{i,j})  
-  S(B_{i,j}\cap B_{i-1,j}) - S(B_{i,j}\cap B_{i,j-1}) \\
&+ S(B_{i,j}\cap B_{i,j-1}\cap B_{i-1,j}\cap B_{i-1,j-1}) 
\Bigg] \,\nonumber
\end{align}
where we have discretized the family of balls on a two-dimensional lattice $B(\sigma) \rightarrow B_{i,j}$. This is based on the following lemma ensuring that this quantity obeys our first criterion.

\paragraph{Lemma:}\label{sec:claim} For a sufficiently fine discretization, Equation \eqref{eqn:diff_entropy} has the same divergences as the envelope of  extremal surfaces anchored to the balls $\{B_{i,j}\}$.

\paragraph{Proof:} First note that divergences in the area of a surface in an asymptotically $AdS$ geometry come from where the surface approaches the asymptotic boundary. 
Since the envelope of a finite family of minimal surfaces is always locally part of one of the surfaces in the family, the divergence of the area of the envelope is supported on the boundary of the union of balls $\partial \left( \bigcup_{i,j} B_{i,j} \right)$. Indeed, in  what follows we will argue that the boundaries of the regions in \eqref{eqn:diff_entropy} work out such that all the divergences cancel for terms in the interior of the sum leaving only contributions from the boundaries of the sum. 

The sum in \eqref{eqn:diff_entropy} contains three types of terms: single balls, and intersections of 2 and 4 balls. Start with the single ball terms, which we will call 1-intersection terms.  The divergence of entanglement entropy of region is an integral of a local quantity on the boundary of the region \cite{hep-th/9901021}. Thus the  divergences from the 1-intersections terms are localized on $\partial B_{i,j}$
\begin{align}
S(B_{i,j}) = S_\mathrm{div} (\partial B_{i,j}) + \mathrm{finite} \,.
\end{align}
Next, we move on to the 2-intersection terms. The intersection of two balls  is bounded by the red curve in the following diagram (dashed lines are boundaries of the individual balls):
\begin{align}
\cdiag{
\draw[dashed,thin]
	(c) circle (1cm) 
	(cr) circle (1cm);
\begin{scope}
\clip (c) circle (1cm);
\draw[red,thick]
	(cr) circle (1cm);
\end{scope}
\begin{scope}
\clip (cr) circle (1cm);
\draw[red,thick]
	(c) circle (1cm);
\end{scope}
}
\end{align}

Families of balls which are tangent to a bulk surface have the property that the boundaries of nearby balls intersect at two points. Suppose $B_{a,b}$ are two neighbouring balls in our discretization of this family. For a sufficiently fine discretization, their centers will be sufficiently close such that each center remains inside the neighbouring ball. If the boundaries of neighboring balls do not cross, this means that one ball is contained inside the other and therefore its bulk extremal surface is strictly closer to the boundary than the other. Therefore this extremal surface cannot touch the envelope. We will only consider ``nice'' families of balls where this does not occur. The condition required to ensure this is given in \eqref{eqn:norm_R_condition} once we have developed the required machinery to address this question.

Since the boundaries intersect at two points, we know that $\partial (B_a \cap B_b)$ will consist of two arcs each following the boundaries of one of the balls. These arcs are of the form $(\partial B_a) \cap B_b$. The divergences will be localised on these arcs as well as at the corners where these arcs meet. In pictures, the boundary of the 2-intersection in red can be separated into the two arcs drawn in red as well as the contributions from two corners given in blue:
\begin{align}
\cdiag{
\draw[dashed,thin]
	(c) circle (1cm) 
	(cr) circle (1cm);
\begin{scope}
\clip (c) circle (1cm);
\draw[red,thick]
	(cr) circle (1cm);
\end{scope}
\begin{scope}
\clip (cr) circle (1cm);
\draw[red,thick]
	(c) circle (1cm);
\end{scope}
}
=
\cdiag{
\draw[dashed,thin,use as bounding box]  (c) circle (1cm) ;
\begin{scope}
\clip (cr) circle (1cm);
\draw[red,very thick] (c) circle (1cm) ;
\end{scope}
}
+
\cdiag{
\draw[dashed,thin,use as bounding box]  (cr) circle (1cm) ;
\begin{scope}
\clip (c) circle (1cm);
\draw[red,very thick] (cr) circle (1cm) ;
\end{scope}
}
+
\cdiag{
\begin{scope}
\clip [name path=A] (c) circle (1cm);
\clip [name path=B] (cr) circle (1cm);
\path [name intersections={of=A and B}] ;
\begin{scope}
\clip (intersection-1) circle (.2cm);
\draw[blue,thick]
	(c) circle (1cm-.4pt) 
	(cr) circle (1cm-.4pt);
\end{scope}
\begin{scope}
\clip (intersection-2) circle (.2cm);
\draw[blue,thick]
	(c) circle (1cm-.4pt) 
	(cr) circle (1cm-.4pt);
\end{scope}
\end{scope}
\pgfresetboundingbox;
\useasboundingbox(.15,-1)(.6,1);
}
\end{align}
In other words, 
\begin{align}
S(B_{i,j} \cap B_{i-1,j}) 
&\sim S_\mathrm{div} (B_{i,j} \cap \partial B_{i-1,j})  
+  S_\mathrm{div} (B_{i-1,j} \cap \partial B_{i,j})  
\\ &+ S_\mathrm{corner} (\partial B_{i,j} \cap \partial B_{i-1,j})^\uparrow
+  S_\mathrm{corner} (\partial B_{i,j} \cap \partial B_{i-1,j})^\downarrow \,,
\end{align}
where $\sim$ indicates that we are dropping finite terms and the notation $(\partial B_{i,j} \cap \partial B_{i-1,j})^\uparrow$ is used to denote the upper corner.

By shifting the index in the sum
\begin{align}
S_\mathrm{div} (B_{i,j} \cap \partial B_{i-1,j}) \rightarrow S_\mathrm{div} (B_{i+1,j} \cap \partial B_{i,j}) \,.
\end{align}
all the divergences can be localised on $\partial B_{i,j}$. To simplify notation, when discussing a fixed ball $B_{i,j}$, we will use $B_\rightarrow$ to denote the ball to the right, ie: $B_{i+1,j}$, and similarly for $B_{\leftarrow,\uparrow,\downarrow}$.
 
Ignoring boundary terms coming from the ends of the sum, the 2-intersection terms can be written 
\begin{align}
\sum_{i,j}  S(B_{i,j}\cap B_{i-1,j}) &\sim 
\sum_{i,j} S_\mathrm{div} (B_\rightarrow \cap \partial B_{i,j})  
+  S_\mathrm{div} (B_\leftarrow \cap \partial B_{i,j}) \label{eqn:lr_div}
\\&+ S_\mathrm{corner} (\partial B_{i,j} \cap \partial B_\leftarrow)^\uparrow
+ S_\mathrm{corner} (\partial B_{i,j} \cap \partial B_\leftarrow)^\downarrow \,. 
\label{eqn:2_intersection_left}
\end{align}
Similar manipulations can be made to the other 2-intersection term to find
\begin{align}
\sum_{i,j}  S(B_{i,j}\cap B_{i,j-1}) &\sim 
\sum_{i,j} S_\mathrm{div} (B_\downarrow \cap \partial B_{i,j})  
+  S_\mathrm{div} (B_\uparrow \cap \partial B_{i,j}) \label{eqn:ud_div}
\\&+ S_\mathrm{corner} (\partial B_{i,j} \cap \partial B_\downarrow)^\rightarrow
+ S_\mathrm{corner} (\partial B_{i,j} \cap \partial B_\downarrow)^\leftarrow \,.
\label{eqn:2_intersection_down}
\end{align}

The 2-intersection terms will cancel the divergences in the 1-intersection term, since for fine enough discretizations 
\begin{align}
\partial B_{i,j} \subset \bigcup_{a\in \{\rightarrow,\leftarrow,\uparrow,\downarrow\} } B_a \,.
\end{align}
However, they will in fact cancel off some regions twice, since the arcs $B_\rightarrow \cap \partial B_{i,j}$ and $B_\uparrow \cap \partial B_{i,j}$ overlap. In other words, $B_\rightarrow \cap B_\uparrow \cap \partial B_{i,j} \neq \emptyset$. The following diagram depicts these facts: the dashed line is $\partial B_{i,j}$, which we can see is covered by  the 4 overlapping arcs where the divergences from \eqref{eqn:lr_div} (in red) and \eqref{eqn:ud_div} (in blue) are localised 
\begin{align}
\cdiag{
\draw [thin,dashed,use as bounding box] (c) circle (1cm);
\begin{scope}
\clip (cr) circle (1cm);
\draw [thick,red] (c)  circle (1cm-.8pt);
\end{scope}
\begin{scope}
\clip (cl) circle (1cm);
\draw [thick,red] (c)  circle (1cm-.8pt);
\end{scope}
\begin{scope}
\clip (cu) circle (1cm);
\draw [thick,blue] (c)  circle (1cm+.8pt);
\end{scope}
\begin{scope}
\clip (cd) circle (1cm);
\draw [thick,blue] (c)  circle (1cm+.8pt);
\end{scope}
}
\end{align}
Cancelling off these regions which have been subtracted twice is where the 4-intersections come in. To be more precise:
\begin{align}
\sum_{i,j} \Big[  &S(B_{i,j}) -  S(B_{i,j}\cap B_{i-1,j}) - S(B_{i,j}\cap B_{i,j-1}) \Big]
\\&\sim \sum_{i,j} \Big[ S_\mathrm{div} (\partial B_{i,j}) 
- S_\mathrm{div} (B_\rightarrow \cap \partial B_{i,j})  
-  S_\mathrm{div} (B_\leftarrow \cap \partial B_{i,j}) 
\\&\qquad\qquad -  S_\mathrm{div} (B_\downarrow \cap \partial B_{i,j})  
-  S_\mathrm{div} (B_\uparrow \cap \partial B_{i,j})
- \mathrm{corners}  \Big]  \,,
\\&\sim \sum_{i,j} \Big[ 
- S_\mathrm{div} (B_\rightarrow \cap B_\downarrow \cap \partial B_{i,j})  
- S_\mathrm{div} (B_\downarrow \cap B_\leftarrow \cap \partial B_{i,j})  
\label{eqn:dbl_subs}
\\&\qquad \qquad 
- S_\mathrm{div} (B_\leftarrow \cap B_\uparrow \cap \partial B_{i,j})  
- S_\mathrm{div} (B_\uparrow \cap B_\rightarrow \cap \partial B_{i,j})  
-\mathrm{corners} \Big] \,.
\end{align}

Finally, we have the 4-intersection term:
\begin{align}
\cdiag{
\draw[dashed,thin] (c) circle (1cm);
\draw[dashed,thin] (cr) circle (1cm);
\draw[dashed,thin] (cu) circle (1cm);
\draw[dashed,thin] (cru) circle (1cm);
\begin{scope}
\clip (c) circle (1cm);
\clip (cr) circle (1cm);
\clip (cu) circle (1cm);
\draw[red,thick] (cru) circle (1cm);
\end{scope}
\begin{scope}
\clip (c) circle (1cm);
\clip (cr) circle (1cm);
\clip (cru) circle (1cm);
\draw[red,thick] (cu) circle (1cm);
\end{scope}
\begin{scope}
\clip (c) circle (1cm);
\clip (cru) circle (1cm);
\clip (cu) circle (1cm);
\draw[red,thick] (cr) circle (1cm);
\end{scope}
\begin{scope}
\clip (cr) circle (1cm);
\clip (cu) circle (1cm);
\clip (cru) circle (1cm);
\draw[red,thick] (c) circle (1cm);
\end{scope}
}
&=
\cdiag{
\coordinate (ca) at (0,0);
\coordinate (cb) at (0.75,0);
\coordinate (cc) at (0,0.75);
\coordinate (cd) at (0.75,0.75);
\draw[dashed,thin,use as bounding box]  (ca) circle (1cm) ;
\begin{scope}
\clip (cb) circle (1cm);
\clip (cc) circle (1cm);
\clip (cd) circle (1cm);
\draw[red,very thick] (ca) circle (1cm) ;
\end{scope}
}
+
\cdiag{
\coordinate (ca) at (0,0);
\coordinate (cb) at (0.75,0);
\coordinate (cc) at (0,0.75);
\coordinate (cd) at (0.75,0.75);
\draw[dashed,thin,use as bounding box]  (cb) circle (1cm) ;
\begin{scope}
\clip (ca) circle (1cm);
\clip (cc) circle (1cm);
\clip (cd) circle (1cm);
\draw[red,very thick] (cb) circle (1cm) ;
\end{scope}
}
+
\cdiag{
\coordinate (ca) at (0,0);
\coordinate (cb) at (0.75,0);
\coordinate (cc) at (0,0.75);
\coordinate (cd) at (0.75,0.75);
\draw[dashed,thin,use as bounding box]  (cc) circle (1cm) ;
\begin{scope}
\clip (ca) circle (1cm);
\clip (cb) circle (1cm);
\clip (cd) circle (1cm);
\draw[red,very thick] (cc) circle (1cm) ;
\end{scope}
}
+
\cdiag{
\coordinate (ca) at (0,0);
\coordinate (cb) at (0.75,0);
\coordinate (cc) at (0,0.75);
\coordinate (cd) at (0.75,0.75);
\draw[dashed,thin,use as bounding box]  (cd) circle (1cm) ;
\begin{scope}
\clip (ca) circle (1cm);
\clip (cb) circle (1cm);
\clip (cc) circle (1cm);
\draw[red,very thick] (cd) circle (1cm) ;
\end{scope}
}
\\&\qquad \qquad+
\cdiag{
\coordinate (ca) at (0,0);
\coordinate (cb) at (0.75,0);
\coordinate (cc) at (0,0.75);
\coordinate (cd) at (0.75,0.75);
\draw[dashed,thin] (ca) circle (1cm);
\draw[dashed,thin] (cb) circle (1cm);
\draw[dashed,thin] (cc) circle (1cm);
\draw[dashed,thin] (cd) circle (1cm);
\begin{scope}
\clip [name path=A] (ca) circle (1cm);
\clip [name path=B] (cb) circle (1cm);
\clip [name path=C] (cc) circle (1cm);
\clip [name path=D] (cd) circle (1cm);
\path [name intersections={of=A and B, name=AB}] ;
\path [name intersections={of=A and C, name=AC}] ;
\path [name intersections={of=B and D, name=BD}] ;
\path [name intersections={of=C and D, name=CD}] ;
\begin{scope}
\clip (AB-1) circle (.2cm);
\draw[blue,thick]
	(ca) circle (1cm-.4pt) 
	(cb) circle (1cm-.4pt);
\end{scope}
\begin{scope}
\clip (AC-1) circle (.2cm);
\draw[blue,thick]
	(ca) circle (1cm-.4pt) 
	(cc) circle (1cm-.4pt);
\end{scope}
\begin{scope}
\clip (BD-2) circle (.2cm);
\draw[blue,thick]
	(cb) circle (1cm-.4pt) 
	(cd) circle (1cm-.4pt);
\end{scope}
\begin{scope}
\clip (CD-2) circle (.2cm);
\draw[blue,thick]
	(cc) circle (1cm-.4pt) 
	(cd) circle (1cm-.4pt);
\end{scope}
\end{scope}
}
\end{align}
The boundary of the 4-intersection is composed of 4 arcs, lying on the boundaries of the 4 balls involved. Notice that the arc lying on $\partial B_{i,j}$ is entirely contained in $B_{i-1,j-1}$ so that 
\begin{align}\label{eqn:4-intersection_sec}
B_{\leftarrow + \downarrow}\cap B_\leftarrow \cap B_\downarrow \cap \partial B_{i,j} = B_\leftarrow \cap B_\downarrow \cap \partial B_{i,j}\,.
\end{align}
This ensures that the contribution from this divergence is exactly right to cancel the regions which were subtracted twice in the 2-intersection case. By shifting the sum index, we find exactly the right terms to cancel the divergences in \eqref{eqn:dbl_subs}. The only thing left to understand is the corners. 

Notice that the corners in the 4-intersection always come from the intersection of only two of the circles. These are the same corners that appeared in the 2-intersection terms. 
\begin{align}
 \cdiag{
\draw[dashed,thin,red] (c) circle (1cm);
\draw[dashed,thin,red] (cr) circle (1cm);
\draw[dashed,dashed] (cu) circle (1cm);
\draw[dashed,dashed] (cru) circle (1cm);
\begin{scope}
\clip [name path=A] (c) circle (1cm);
\clip [name path=B] (cr) circle (1cm);
\clip [name path=C] (cu) circle (1cm);
\clip [name path=D] (cru) circle (1cm);
\path [name intersections={of=A and B, name=AB}] ;
\begin{scope}
\clip (AB-1) circle (.2cm);
\draw[blue,thick]
	(c) circle (1cm-.4pt) 
	(cr) circle (1cm-.4pt);
\end{scope}
\end{scope} 
 }
 \rightarrow
 \cdiag{
\draw[dashed,thin,red] (c) circle (1cm);
\draw[dashed,thin,red] (cr) circle (1cm);
\begin{scope}
\clip [name path=A] (c) circle (1cm);
\clip [name path=B] (cr) circle (1cm);
\path [name intersections={of=A and B, name=AB}] ;
\begin{scope}
\clip (AB-1) circle (.2cm);
\draw[blue,thick]
	(c) circle (1cm-.4pt) 
	(cr) circle (1cm-.4pt);
\end{scope}
\end{scope} 
 }
 \end{align}
 The four corners are $(\partial B_{i,j} \cap \partial B_\downarrow)^\leftarrow$, 
 $(\partial B_{i,j} \cap \partial B_\leftarrow)^\downarrow$, 
 $(\partial B_{i-1,j} \cap \partial B_\downarrow)^\rightarrow$ and 
 $(\partial B_{i,j-1} \cap \partial B_\leftarrow)^\uparrow$. By shifting the index in the last two corners, we recover the corner contributions from \eqref{eqn:2_intersection_left} and \eqref{eqn:2_intersection_down} with the opposite sign so that they cancel exactly. This completes the proof of the lemma.

In this argument, we have used that for fine enough discretizations:
\begin{align}
 \partial B_{i,j} \subset \bigcup_{a\in \{\rightarrow,\leftarrow,\uparrow,\downarrow\} } &B_a \,,  \label{eqn:ball_facts1} \\
 \Big(B_\rightarrow \cap \partial B_{i,j} \Big) \cap \Big( B_\uparrow \cap \partial B_{i,j} \Big) &\neq \emptyset \,, \label{eqn:ball_facts}\\
 B_\rightarrow \cap B_\uparrow \cap \partial B_{i,j} &\subset B_{\rightarrow + \uparrow} \,. \label{eqn:ball_facts3}
\end{align}
We demonstrate \eqref{eqn:ball_facts1}-\eqref{eqn:ball_facts3} in Appendix~\ref{app:discretization}.

\paragraph{Higher dimensional generalisations}
We have restricted ourselves to $d=2$ (i.e., $2+1$-dimensional field theories) in this work to simplify the notation and to make arguments simpler. However we do not see any obstruction to extending this argument to higher dimensions with an additional term in the alternating sum for each additional dimension. 

In our analysis for $d=2$, we saw that the 1-intersection terms had divergences localised on 1-dimensional curves. The 2-intersection terms had new divergences localised on 0-dimensional points at the corners of where these curves intersect. However, the 4-intersection terms did not introduce any new divergences and so were able to cancel all the divergences appearing in the previous terms.

In higher dimensions, 1-intersection terms have divergences localised on a $(d-1)$-dimensional surface. 2-intersection terms involves additional divergences at the corners where the boundaries of the balls intersect on a $(d-2)$-dimensional surface. Now 4-intersection terms, formed by the intersection of two regions, each themselves the intersection of two balls, will introduce new divergences where the corners meet on a $(d-3)$-dimensional surface.  $2^{k-1}$-intersection terms, formed by taking the intersection of two $2^{k-2}$-intersection regions, introduce new divergences on $(d-k)$-dimensional surfaces. 
This will continue for the first $d$ terms, the last of which will have divergences localised on a $0$-dimensional surface, that is on isolated points. However, the $(d+1)^\mathrm{th}$ term, a $2^d$-intersection term, cannot introduce new divergences since the intersections of isolated points are again points, or in other words there are no $(-1)$-dimensional surfaces. 

In summary, each alternating term in the sum introduces divergences on regions with a higher co-dimension supported on the intersections of intersections. However if there are $d+1$  alternating terms, the last term will not introduce a new type of divergence and so can at least in principle cancel off the remaining divergences from the other terms. We therefore conjecture that the general formula is
\begin{align}\label{eqn:high_dim_diff_ent_discretized}
S_\mathrm{diff}&\big( \{ B_I \} \big)=
\sum_{I = (i_1,\ldots, i_d)} \Big[ S\big(B_I\big) 
-\sum_{n=1}^d S\big(B_I \cap B_{I-e_n}\big) \\
&+ \sum_{n \neq m} S\big(B_I \cap B_{I-e_n} \cap B_{I-e_m} \cap B_{I-e_n-e_m} \big) 
\\& - \sum_{n \neq m \neq l} S\big(B_I 
\cap B_{I-e_n} \cap B_{I-e_m} \cap B_{I-e_l} 
\cap B_{I-e_n-e_m} \cap B_{I-e_n-e_l}
\\&\qquad\qquad \qquad\qquad \cap B_{I-e_m-e_l}  
\cap B_{I-e_n-e_m-e_l} \big)  + \ldots \Big] \,
\end{align}
where $I$ is a multi-index indexing the discretized family of balls and $e_n$ is a unit vector shifting the $n^\mathrm{th}$ index of this multi-index.

\subsection{Continuum limit}
In this section we will find the continuum limit of \eqref{eqn:diff_entropy}. First, note that while the intersection of two nearby intervals is again an interval, in higher dimensions the intersection of two balls is not a ball. However, the intersection of two nearby balls is nearly a ball. The correct generalisation of $\partial_v S(u,v(u))$ appearing in the differential entropy formula is not to find some differential operator like $\partial_R$ acting on $S(x_0^{(i)},R(x_0^{(i)}))$, but instead to recognise $\partial_v S(u,v(u))$ in the two-dimensional differential entropy as a derivative by a deformation of the shape of the interval.  So the correct generalisation to higher dimensions involves shape derivatives of the entanglement entropy.\footnote{Shape derivatives of entanglement entropy have been considered by \cite{1304.7100,1411.7011,1511.05179}.}

To be precise, we can think of the entanglement entropy as a functional of the entangling region in the CFT.  Specifically, $S(A)$ for a region $A$ is a functional of $\partial A$ since a region is defined by its boundary. A simple region can be described by giving its boundary in polar coordinates. For example, the boundary of a ball $(R,x_0^{i})$ is given by
\begin{align}\label{eqn:incl_map_R_K}
 r(\theta | R,x_0^i) = \sqrt{\big(x_0^{(1)} + R \cos \theta\big)^2+\big(x_0^{(2)} + R \sin\theta\big)^2} \,.
 \end{align}
Our polar coordinates can alternatively be parametrized by a unit vector $\Omega^i=(\cos\theta,\sin\theta)$, so that
\begin{align}
r(\Omega^i | R,x_0^i) &= \big|R \Omega^i +  x_0^i \big| \,.
\end{align}
This gives a parametrization of the space of regions, which we will denote $\mathcal{R}$. 
We can then define shape derivatives by thinking of the entanglement entropy as a functional on $\cR$, $S[r(\theta)]$. 
The inclusion map from the space of balls $\cK$ to the space of regions in this parameterization, $\cK \rightarrow \cR$, is given by \eqref{eqn:incl_map_R_K}. By taking derivatives of that expression we can understand the shape deformation that corresponds to moving around in the space of balls, which can be given explicitly as
\begin{align}
\dot R \partial_R + \dot x_0^i \partial_{x_0^i} :\rightarrow \delta r(\Omega^i|B) = \big( \dot R \Omega^i + \dot x_0^i \big)   \frac{\big(  R \Omega^i +  x_0^i \big) }{\big|R \Omega^k +  x_0^k \big| } \,,\label{eqn:explicit_dr}\\
\delta r (\theta | R=1,x_0=\{0\}) = \dot R + \dot x_0^{(1)} \cos\theta + \dot x_0^{(2)} \sin \theta \,.
\end{align}
Namely these deformations change one ball into a neighboring ball.

Now we can address the conditions required to ensure that 
none of the balls in the family are contained in a neighbouring ball (as discussed earlier this condition  ensures that  the extremal bulk surface attached to every ball is tangent to the envelope of extremal surfaces). Without loss of generality, consider the ball at the origin ($x_0^i = \{0\}$). From the perspective of shape deformations, this ball is contained in (contains) its neighbour if the shape deformation between these balls is strictly negative (positive) for all angles $\Omega^i$. Therefore, to ensure that this does not occur, every shape deformation---in any direction along the family of balls---must have zeros.

Consider moving the center of the ball away from the origin in the direction $V^i$ in the boundary field theory. In the space of balls $\cK$ the vector pointing in this direction is
\begin{align}
V^i \frac{\partial R}{\partial x_0^i} \partial_R + V^i \partial_{x_0^i} \,.
\end{align}
Plugging this vector into \eqref{eqn:explicit_dr}, the corresponding shape deformation is
\begin{align}
\delta r(\Omega) = \big(\frac{\partial R}{\partial x_0^i} +  \Omega_i \big)    V^i    \,.
\end{align}
Since $\Omega_i$ is an arbitrary unit vector, $\Omega \cdot  V$ takes every value in 
\begin{align}
-|V| \leq \Omega \cdot V \leq |V| \,.
\end{align}
The condition
\begin{align}\label{eqn:norm_R_condition}
\left|\frac{\partial R}{\partial x_0^i}\right|<1 
\end{align}
ensures that
\begin{align}
\frac{\partial R}{\partial \vec x_0^i}  V^i 
\leq  \left| \frac{\partial R}{\partial  x_0^i} \right|  |V|
\leq |V|
\end{align}
so that the shape deformation for any $V^i$ has a zero for some $\Omega^i$ and therefore does not have a definite sign.

On the other hand, if the condition is not satisfied, there is some $V_*$, aligned with $\frac{\partial R}{\partial \vec x_0^i} $, so that 
\begin{align}
\frac{\partial R}{\partial  x_0^i}  V_*^i = \left| \frac{\partial R}{\partial  x_0^i} \right|  |V_* | \geq  |V_*| \,
\end{align}
which ensures that the shape deformation has a definite sign.  Thus, we will define a ``nice'' family of balls to be one which obeys the condition \eqref{eqn:norm_R_condition}.

A causal structure can be defined on the space of balls $\cK$ by defining a ball to be in the future of another if it is entirely contained in it \cite{1505.05515}. 
In the language of this causal structure, a family of balls where none of the balls in the family is contained in a neighbouring ball, i.e., what we called a ``nice'' family of balls, is a surface in $\cK$ where none of the tangents are timelike. In other words, a family of balls is nice if and only if it is a spacelike co-dimension 1 surface in $\cK$.\footnote{
In fact, the explicit form of the shape deformation given in \eqref{eqn:explicit_dr} tells us that neighbouring balls are timelike, null or spacelike if and only if 
\begin{align}
-\Delta R^2 + (\Delta x_0^i)^2
\end{align}
is negative, zero or positive respectively. 
Thus, any metric on $\cK$ respecting this notion of causality must have the form 
\begin{align}
ds^2_\cK = e^{2\omega} \Big( -dR^2 + (dx_0^i)^2  \Big)
\end{align}
for some conformal factor $e^{2\omega}$. This metric is consistent with that found in \cite{1505.05515,1606.03307}.
}

\subsubsection{An explicit expression for shape deformations in embedding space}
\label{sec:deformations_embedding}
The area of a bulk surface cannot depend on our choice of conformal frame on the boundary. In order to make explicit how quantities transform it is useful to write them in embedding space.  Thus, we seek a formula for differential entropy that is built out of shape derivatives of entanglement entropy, and is independent of the conformal frame.  We will denote both the ball and the embedding space vector representing it by $B$.  In embedding space, points in the boundary theory lift to null rays, denoted $Z$. The points inside a ball $B$ are those with $Z\cdot B >0$; therefore the boundary of the ball satisfies $Z\cdot B=0$. In $d=2$, in analogy to the way we described the boundary of a ball  using polar coordinates $r(\theta|R,x_0^i)$, the rays $Z$ orthogonal to $B$ can be parametrized by an angular coordinate: $Z(\theta | R,x_0^i)$. In general dimensions we will need multiple angles to parametrize the boundary, so we will use the notation $Z(\Omega | B)$, where $\Omega$ is a unit vector on the $(d-1)$-dimensional sphere.

We will now describe small deformations of the ball $B$. Deformations correspond to slightly deforming the points on the boundary of the ball, by an amount $\delta Z$ to a new boundary point $Z'$, 
\begin{align}
Z'(\Omega) = Z(\Omega | B) + \delta Z(\Omega)\,.
\end{align}
This new point $Z'$ is still parametrized by a null ray, so $Z'^2=0$ and therefore $\delta Z \cdot Z=0$. Since points in the CFT correspond to rays in embedding space, deformations proportional to $Z$ have no effect and can be dropped. The deformations proportional to $\frac{\partial Z}{\partial \Omega}$ will correspond to reparametrizations of the ball which do not change the actual shape being described. There will be $d-2$ of these, leaving us with a one dimensional family of non-trivial deformations. As $\frac{\partial Z}{\partial \Omega}$ and $Z$ are orthogonal to $B$, the only non-trivial possibility is that $\delta Z$ is in the direction of $B$. So deformations can be parametrized by a single scalar function 
\begin{align} \label{eqn:embed_rho_def}
\delta Z(\Omega) = -\rho(\Omega) B\,. 
\end{align}
Since the points inside the ball obey $Z\cdot B>0$, deformations towards the outside (inside) of the ball are those with $\rho$ positive (negative). Thus $\rho$ describes the normal component of the shape deformation.

Let us look at the example of $B=(0,0,0,1)$, whose boundary is the unit circle at the origin ($x^2+y^2=1$):
\begin{align}
Z(\theta | 1,\{0\}) = (1,\cos\theta,\sin\theta,0)\,, \qquad
\frac{\partial Z}{\partial \theta}=(0,-\sin \theta, \cos\theta,0)\,.
\end{align}
In polar coordinates, this circle is $r(\theta)=1$. Deformations of the circle are parametrized by a function $\delta r(\theta)$, so that deformed points are at $x=(1+\delta r) \cos\theta, y= (1+\delta r)\sin\theta$. In embedding space these points are 
\begin{align}
Z' = ((1+\delta r), (1+\delta r) \cos\theta, (1+\delta r) \sin\theta, - \delta r ) + O(\delta r^2).
\end{align}
Since boundary points are described by null rays, $Z'$ can be rescaled by $(1+\delta r)^{-1}$ and we see explicitly that to leading order in $\delta r$, 
\begin{align}
Z'(\theta) = Z(\theta) - \delta r(\theta) B. \label{eqn:lambda_dr_relation}
\end{align}
In this simple case, $\rho$ is exactly $\delta r$.

The deformation that moves a ball towards another is simple to derive in this language (this is another description of the tangent space to the embedding of the space of balls $\cK$ in the space of regions $\cR$). If the new ball is $B' = B + \delta B$, then $Z' \cdot B'=0$, so that $\rho =  -\delta Z \cdot B = \delta B \cdot Z$. By taking derivatives of our explicit parametrization of $B$ we can reproduce \eqref{eqn:explicit_dr}. This form makes the zeros of the shape deformation easy to find: they occur when $Z$ is orthogonal to $\delta B$.

 Given a small deformation of a ball $B$, parametrized as described above by $\rho(\theta)$, we can denote functional derivatives of the entanglement entropy around particular ball shaped region $B$ by\footnote{Functional derivatives are analogous to those familiar from vector calculus. To make our notation clearer we have written below the vector calculus version of these expressions. Given a function $F(x_i)$ of $n$ variables, $i=1 \ldots n$,
 \begin{align}
\delta^{(1)} F[v_i] &= \sum_{i=1}^n \frac{\partial F}{\partial x_i} v_i \,,\\
\delta^{(2)} F[v_i,w_j] &= \sum_{i,j=1}^n \frac{\partial F}{\partial x_i \partial x_j} v_i w_j \,.
\end{align}
We have suppressed the dependence on the point $\{ x_i \}$ where these derivatives are evaluated. In our analogy, this is the dependence on which ball we perturb around.
 }
\begin{align}\label{eqn:shape_deriv_def}
\delta^{(1)} S [\rho ] &= \int d\theta \frac{\partial S}{\partial \delta r(\theta)} \rho(\theta) \,,\\
\delta^{(2)} S [\rho_1, \rho_2] &= \int d\theta_1 d\theta_2 \frac{\partial^2 S}{\partial \delta r(\theta_1) \partial \delta r(\theta_2)} \rho_1 (\theta_1) \rho_2(\theta_2) \,.
\end{align}

\subsubsection{Discretizations that respect the continuum symmetries} \label{sec:discretizations}
We will arrive at our formula for differential entropy by starting with the discretized version in \eqref{eqn:diff_entropy}, and then taking a continuum limit to get an expression in terms of shape derivatives.   Families of discretizations which have a nice continuum limit can be constructed by choosing coordinates on our family of balls in $\cK$ and putting down an integer lattice with spacings $a$ in those coordinates. Instead of picking specific coordinates we will keep track of this freedom, since the requirement that our final answer be independent of our choice of coordinates will provide an important constraint. Given coordinates $\sigma: \mathbb{R}^{d} \rightarrow \cK$, the discretization is $B_{i,j} = \sigma(a i,a j)$ for $i,j\in\mathbb{Z}$.

Consider the shape of $B_{i,j} \cap B_{i-1,j}$ appearing in \eqref{eqn:diff_entropy}. The boundary of this region follows the boundaries of the regions $B_{i,j}$ and $B_{i-1,j}$. Think of the boundary of $B_{i-1,j}$ as a deformation of the boundary of $B_{i,j}$. In other words, expand $Z(\Omega| B_{i-1,j})$ around $Z(\Omega| B_{i,j})$: 
\begin{align}
Z^{A}(\Omega | B_{i-1,j}) =  Z^{A}(\Omega | B_{i,j}) - \rho^{\leftarrow}(\Omega) B_{i,j}^{A} + O(B_{i,j}-B_{i-1,j})^2, 
\end{align}
where the superscript in $\rho^\leftarrow$ denotes that this is the deformation from moving towards the ball to the left in our discretization.
$\rho^\leftarrow$ can be computed by using the fact that $Z(\Omega | B_{i-1,j}) \cdot B_{i-1,j} =0$ and expanding both sides:
\begin{align}
0&= \left( Z(\Omega; B_{i,j}) - \rho^\leftarrow(\Omega) B_{i,j} \right) \cdot \big( B_{i,j} - (B_{i,j} - B_{i-1,j} ) \big) \,,\\
&\quad \rho^\leftarrow = - \left( B_{i,j} - B_{i-1,j} \right) \cdot Z  \equiv - (\delta B)^\leftarrow \cdot Z \,.
\end{align}
In the continuum limit, $ (\delta B)^\leftarrow \rightarrow \partial_{\sigma^{(1)}} B(\sigma)$.

As discussed below \eqref{eqn:embed_rho_def}, the deformation points inwards when $\rho<0$. Thus the boundary of $B_{i,j} \cap B_{i-1,j}$ will follow $\partial B_{i-1,j}$ exactly when $\rho^\leftarrow$ is negative. 
\begin{align}
\label{diag:inside}
\cdiag{
\draw[dashed,thin,use as bounding box]
	(cr) circle (1cm);
\begin{scope}
\clip (c) circle (1cm);
\draw[red,thick]
	(cr) circle (1cm);
\end{scope}
\begin{scope}
\clip (cr) circle (1cm);
\draw[red,thick]
	(c) circle (1cm);
\end{scope}
}
\end{align}
The boundary of $B_{i,j} \cap B_{i-1,j}$ will correspond to the deformation 
\begin{align}
\lceil \rho^\leftarrow \rceil \equiv H(\rho^{\leftarrow}(\Omega)) \,,
\label{eq:ceilingdef}
\end{align}
where $H$ is the ramp function
\begin{align}\label{eqn:step_def}
H(x) = \begin{cases}
0 &\qquad x\geq0 \\
x &\qquad x<0
\end{cases} \,.
\end{align}

This means that $S(B_{i,j}\cap B_{i-1,j})$ can be written as an expansion
\begin{align}
S(B_{i,j}\cap B_{i-1,j}) = S(&B_{i,j}) 
+ \delta^{(1)} S\big[\lceil \rho^\leftarrow \rceil \big] \\
&+\delta^{(2)} S\big[\lceil \rho^\leftarrow \rceil ,\lceil \rho^\leftarrow \rceil \big] 
+ \delta^{(1)} S\big[(\delta^2 \rho)^\leftarrow \big] +O(a^3) \,,
\end{align}
where $(\delta^2 \rho)^\leftarrow$ is the shape deformation at second order in the lattice spacing $a$ and $\delta^{(1)} S$ and $\delta^{(2)} S$ are the shape derivatives defined in \eqref{eqn:shape_deriv_def}.

Now on to the 4-intersection term:
\begin{align}
\cdiag{
\draw[dashed,thin,use as bounding box]
	(c) circle (1cm);
\begin{scope}
\clip (c) circle (1cm);
\draw[red,dashed]
	(cl) circle (1cm);
\end{scope}
\begin{scope}
\clip (cl) circle (1cm);
\draw[red,dashed]
	(c) circle (1cm);
\end{scope}
\begin{scope}
\clip (c) circle (1cm);
\draw[red,dashed]
	(cd) circle (1cm);
\end{scope}
\begin{scope}
\clip (cd) circle (1cm);
\draw[red,dashed]
	(c) circle (1cm);
\end{scope}
\begin{scope}
\clip (c) circle (1cm);
\clip (cd) circle (1cm);
\clip (cl) circle (1cm);
\draw[blue,thick]
	(cld) circle (1cm-.5pt);
\end{scope}
\begin{scope}
\clip (c) circle (1cm);
\clip (cld) circle (1cm);
\clip (cl) circle (1cm);
\draw[blue,thick]
	(cd) circle (1cm-.5pt);
\end{scope}
\begin{scope}
\clip (c) circle (1cm);
\clip (cd) circle (1cm);
\clip (cld) circle (1cm);
\draw[blue,thick]
	(cl) circle (1cm-.5pt);
\end{scope}
\begin{scope}
\clip (cld) circle (1cm);
\clip (cd) circle (1cm);
\clip (cl) circle (1cm);
\draw[blue,thick]
	(c) circle (1cm-.5pt);
\end{scope}
}
\end{align} 
As discussed above \eqref{eqn:4-intersection_sec}, its boundary consists of 4 arcs following a bit of the boundary of each of the balls involved. The parts following the balls $B^\leftarrow$ and $B^\downarrow$ will follow $\rho^\leftarrow$ and $\rho^\downarrow$ respectively. 
To understand the part following the ball $B^{\leftarrow+\downarrow}$ (i.e., the ball that is diagonally down and to the left), expand the boundary of that ball as
\begin{align}
Z^{A}(\Omega | B_{i-1,j-1}) =  Z^{A}(\Omega | B_{i,j}) -\Big( \rho^{\leftarrow}_{i,j}(\Omega) +\rho^{\downarrow}_{i,j}(\Omega) \Big)B_{i,j}^{A}  + O(a^2) \,.
\end{align}
To linear order, the shape deformation is just the sum of the deformations in each direction
\begin{align}
\rho^{\leftarrow+\downarrow} = \rho^{\leftarrow} + \rho^{\downarrow} \,. 
\end{align}
Since the ramp function is additive, we find that at linear order the combined shape deformation that describes moving towards the 4-intersection term is simply
\begin{align}
\lceil \rho^{\leftarrow} \rceil+ \lceil\rho^{\downarrow}\rceil \,.
\end{align}
As this is the sum of two piecewise functions, there will be 4 domains corresponding to the 4 arcs of the boundary of the 4-intersection region.

Expanding the entropy of the 4-intersection term we find
\begin{align}
S(B_{i,j} \cap &B_{i-1,j} \cap B_{i,j-1} \cap B_{i-1,j-1}) = S(B_{i,j})
+ \delta^{(1)}S\big[ \lceil \rho^{\leftarrow} \rceil+ \lceil\rho^{\downarrow}\rceil\big] \\
&+ \delta^{(2)}S\big[ \lceil \rho^{\leftarrow} \rceil+ \lceil\rho^{\downarrow}\rceil
,  \lceil \rho^{\leftarrow} \rceil+ \lceil\rho^{\downarrow}\rceil \big] 
+ \delta^{(1)}S[(\delta^2\rho)^{\leftarrow+\downarrow}] +O(a^3) \,.
\end{align}
Since $\delta^{(1)}S$ and $\delta^{(2)}S$ are linear and bi-linear respectively in their arguments, many of the terms in this expression will cancel with the 2-intersection terms which come with the opposite sign. In particular, all the terms linear in $a$ will cancel and the resulting expression is $O(a^2)$:
\begin{align}
S_{diff}\Big[\{B_{i,j}\}\Big] = 
\sum_{i,j} \Bigg[ 2 \delta^{(2)}S\Big[ \lceil \rho^{\leftarrow} \rceil,
				\lceil\rho^{\downarrow}\rceil \Big]
+  \delta^{(1)}S \Big[ (\delta^2\rho)^{\leftarrow+\downarrow}-(\delta^2\rho)^{\leftarrow}-(\delta^2\rho)^{\downarrow} \Big] +O(a^3)\Bigg] \,,
\end{align}
where the shape derivatives $\delta^{(1)}S$ and $\delta^{(2)}S$ are defined explicitly in \eqref{eqn:shape_deriv_def}.

The entanglement entropy of a region in field theory is divergent, but it has a universal piece which is regulator independent. In odd dimensions, this universal piece is the finite part of the entanglement entropy. The change to this universal part due to changes of the shape of the region can be extracted to give a regulated version of the shape derivative. The second shape derivative $\frac{\partial^2 S}{\partial \delta r(\theta_1) \partial \delta	r(\theta_2)}$ includes a non-local contribution which effects the universal part and contributes to the regulated shape derivative as well as local parts, proportional to $\delta(\theta_1-\theta_2)$, which are divergent. $\frac{\partial S}{\partial \delta r(\theta)}$ on the other hand only contributes to the divergences.\footnote{There are two ways of seeing this. First, there are no non-local terms we can write for a shape derivative at first order, so this must purely be a divergent contact term. Second, if you try computing this explicitly for a minimal surface anchored on a ball in $AdS$ you will see this is the case. See \cite{1511.05179} for a thorough discussion of shape derivatives of entanglement entropy and their divergences in field theory.} The Lemma demonstrated in Section \ref{sec:claim} guarantees that the differential entropy is finite, so all these divergent terms must exactly cancel against each other leaving only the regulated second shape derivative,
\begin{align}
S_{diff}\Big[\{B_{i,j}\}\Big] = 2 \sum_{i,j} \delta^{(2)}S_\mathrm{reg}\Big[ \lceil \rho^{\leftarrow} \rceil, \lceil\rho^{\downarrow}\rceil \Big] +O(a^3) \,,
\end{align}
where $\delta^{(2)}S_\mathrm{reg}$ is the regulated second shape derivative, i.e., the second shape derivative of the universal finite part of the entanglement entropy.
This has the form of a Riemann integral and in the continuum limit, the differential entropy becomes an integral of the second shape derivative over the family of balls
\begin{align} \label{eqn:diff_ent_def}
S_{diff}\Big[\{B(\sigma) \}\Big] = \frac{\pi}{2} \int d^2\sigma \mu(\sigma)  \, \delta^{(2)}S_\mathrm{reg} 
\Big[ \lceil \rho_{T^\leftarrow}\rceil , \lceil\rho_{T^\downarrow} \rceil \Big]\,.
\end{align}
In this expression, $\mu$ is a measure on the set of balls defining our bulk surface and $T^{\leftarrow,\downarrow}$ are a non-degenerate pair of vector fields tangent to the family of balls.   We included a factor of $\pi/2$ in the normalization of the measure for later convenience.   We also introduced the new notation $\rho_T$, which is the deformation towards a ball in $\cK$ in the direction of a vector $T$. In embedding space this deformation is simply
\begin{align}
\rho_T (\theta) \equiv T \cdot Z(\theta | B)\,.
\end{align}

The choice of discretization can be seen through the appearance of the measure $\mu$ which keeps track of the density of points and $T^{\leftarrow,\downarrow}$ which keep track of the directions of the difference vectors of the discretization. In terms of the coordinates $\sigma$ that the discretization was based on, the measure $\mu$ is that inherited from the Euclidean metric pushed through the coordinates, and the vectors $T^{\leftarrow,\downarrow}$ are the basis vectors defined by the coordinates.

If we change the coordinates that our discretization is based on, the residual measure and tangent vectors will also change. The change of coordinates $\sigma \rightarrow \sigma'$ has Jacobian $J_{A,B} \equiv \frac{\partial B^A(\sigma'^i)}{\partial B^B(\sigma^j)}$, then 
\begin{align} \label{eqn:param_trans}
\mu \rightarrow \det(J) \mu\,, \qquad T \rightarrow J \cdot T\,,
\end{align}
so that our expressions so far depend on the discretization lattice used to construct it. However, we will argue that the conformal invariance of the theory (not of the state) picks out a special class of discretizations that lead to an unambiguous definition of the differential entropy.

\subsection{Differential entropy as the area of bulk surfaces}
Since the area of a bulk surface must be independent of our choice of conformal frame on the boundary, we should demand the same from our definition of differential entropy.
There is a special class of coordinates adapted to this $SO(d+1,1)$ symmetry: 
the coordinates such that the metric obtained by pushing forward the Euclidean metric on $\mathbb{R}^d$ gives the unique $SO(d+1,1)$ invariant metric on $\cK$ \cite{1606.03307}
\begin{align} 
ds^2 = \left( \sum_k \frac{\partial \sigma^i}{\partial x^k } \frac{\partial \sigma^j}{\partial x^k } \right)^{-1} d\sigma^i d\sigma^j =  dB(\sigma) \cdot dB(\sigma) \,,
\end{align}
where $\sigma^i(x^k)$ are coordinates on our family of balls ($\sigma:\mathbb{R}^d \rightarrow \cK$), $B(\sigma)$ is the embedding space vector associated to the ball $\sigma$, and $\cdot$ is the standard $SO(d+1,1)$ invariant inner product on embedding space.   We will carry out the discretization and limits described above in this class of coordinates.   Since these coordinates lead to the unique $SO(d+1,1)$ invariant metric on our family of balls, the continuum measure $\mu$ that we obtain is the unique $SO(d+1,1)$ invariant measure on our family of balls and the difference vectors $T_{\leftarrow,\downarrow}$, which are always orthogonal in the original Euclidean metric on $\mathbb{R}^d$, will be orthogonal as embedding space vectors. The embedding space vectors $T_{\leftarrow,\downarrow}$ thereby provide an orthonormal frame for the tangent space to our family of balls in embedding space.  Conceptually, this procedure corresponds to placing a regular grid on embedding space. 

In fact, in order for the differential entropy to be $SO(d+1,1)$ invariant, $\mu$ must be the unique $SO(d+1,1)$ invariant measure, since the shape derivative is a $SO(d+1,1)$ scalar.  So if we discretize and take limits in a manner consistent with symmetries, we must produce this measure.    The procedure that we followed above has the additional virtue of unambiguously specifying the vectors $T$ that transport between neighboring balls.  This procedure is analogous to making a gauge choice well adapted to the symmetries of our problem to do explicit computations.


Our prescription for fixing the definition of differential entropy makes sense in any state and so we conjecture that the differential entropy 
computes the area of the bulk envelope associated to the family of balls used to define it:
\begin{align}\label{eqn:area_diff_ent}
\frac{1}{4 G_N} Area(N) =  S_\mathrm{diff} \big[ B(\sigma) \big]\,.
\end{align}
Here $N$ is the envelope of the family of balls $B(\sigma)$ and we have used the standard holographic matching \cite{1405.7862}\footnote{This reference uses $d$ to denote boundary spacetime dimensions, whereas we have used $d$ to denote only spatial dimensions.} where 
\begin{align}
\frac{L_{AdS}^d}{4G_N} = \frac{2 d  \pi^{\frac{d+3}{2}} \Gamma(\frac{d+1}{2}) }{\Gamma(d+3) } C_T  
\xrightarrow{d=2}
\frac{\pi^3}{12} C_T   \,.
\label{eq:CT}
\end{align}
Here $C_T$ is the overall normalization of the two-point function of the stress tensor.

As we discussed in Section~\ref{sec:int_geo}, simply integrating the invariant measure on the space of balls is sufficient to reproduce the area of an envelope in empty AdS, but it does not give any indication of what measure to use when moving away from the vacuum. From this perspective, our conjecture is that the shape derivatives of entanglement entropy provide the necessary state dependence and that \eqref{eqn:area_diff_ent} provides the appropriate measure to use in other states. In other words, although this prescription for defining differential entropy makes use of the conformal symmetry of the theory, it does not assume anything about the state.

Our conjecture is supported by three pieces of evidence: (1) \eqref{eqn:area_diff_ent} has the correct structure of UV divergences, (2) it has the appropriate invariance under changes of conformal frame, and (3) we will verify that it is correct for the vacuum state in the next section.  Proving this conjecture requires a computation of shape derivatives of entanglement entropy in excited states, which would be interesting to carry out in its own right, but has not yet appeared in the literature.\footnote{In principle, the computation of these shape derivatives is not problematic, whether we take a bulk perspective following \cite{1304.7100,1411.7011} or a conformal perturbation theory perspective following \cite{1511.05179}, although the former should be a more straightforward computation. None the less, the need to regulate the answer introduces sufficient technical complications that we will leave this for follow up work.}

\section{Differential entropy computes area in the vacuum}
\label{sec:verify}
In the previous section, we defined a generalisation of differential entropy in terms of second shape derivatives of the entanglement entropy:
\begin{align}
\delta^{(2)}S_\mathrm{reg} 
\Big[ \lceil \rho_{T_1}\rceil , \lceil\rho_{T_2} \rceil \Big]  
= \int d\theta_1 d\theta_2 \,
\frac{\partial^2 S_\mathrm{reg}}{\partial \delta r(\theta_1) \partial \delta r(\theta_2)} 
\lceil \rho_{T_1}\rceil (\theta_1) \lceil\rho_{T_2} \rceil(\theta_2)
\,.
\end{align}
This formula is well defined for any state, but explicit formulae for the shape derivatives of entanglement entropy only appear in the literature for the vacuum state of the CFT, so we will start there. 

The finite part of the second shape derivative in vacuum expressed in Fourier space is given by Mezei's formula \cite{1411.7011,1511.05179}
\begin{align}
\int d\theta_1 d\theta_2 &\frac{\partial^2 S_\mathrm{reg}}{\partial \delta r(\theta_1) \partial \delta r(\theta_2)} e^{i k_1 \theta_1} e^{i k_2 \theta_2} = - \frac{\pi^4}{6} C_T \,\delta_{k_1,-k_2} |k_1| (k_1^2-1)  \,.
 \end{align}
 where $C_T$ was defined in \eqref{eq:CT}.
Since this result has already been appropriately regulated, it will be most convenient for our purposes to work in Fourier space were we can apply it directly.
 
To make use of this result, we need the Fourier transform of our shape deformations
\begin{align}
\lceil \rho_{T_i}\rceil(\theta) = H\big( Z(\theta) \cdot T_i \big) \,,
\end{align}
where $Z(\theta)$ parametrizes the boundary of the ball $B$ around which this shape derivative is to be evaluated and $H$ is the ramp function defined in \eqref{eqn:step_def}. Recall that points on the boundary correspond to null rays, $Z^2=0$ and $Z \sim \lambda Z$. We can express these in a basis adapted to the vectors $T_{1,2}$. First, introduce a timelike vector $W$ orthogonal to both $T_{1,2}$ and $B$, which we can do since in this case the embedding space is four dimensional.  We then use Gram-Schmidt to construct $\tilde T_2$ such that $T_1 \cdot \tilde T_2 =0$ to give an orthonormal basis $\{W,T_1,\tilde T_2,B\}$:
\begin{align}
\tilde T_2 &\equiv \frac{T_2 -(T_1 \cdot T_2) T_1 }{\sqrt{1-(T_1\cdot T_2)^2}} \,. 
\end{align}
The points on $\partial B$ can be expressed in terms of this basis as 
\begin{align}
Z(\theta) = W - \cos\theta\, T_1 - \sin \theta\, \tilde T_2 \,,
\end{align}
where we have used the freedom to rescale $Z$ to set the coefficient of $W$ to $1$. 

In terms of this parametrization,
\begin{align}
Z(\theta)\cdot T_1 &= - \cos\theta \,, \\
Z(\theta) \cdot T_2 &= - (T_1\cdot T_2)  \cos\theta - \sqrt{1 - (T_1 \cdot T_2)^2} \,\sin\theta  \,.
\end{align}
Since $T_{1,2}$ are unit vectors, $T_1 \cdot T_2  = \cos\alpha$, where $\alpha$ is the angle between them. This allow us to re-express
\begin{align}
Z(\theta) \cdot T_2 = - \cos(\theta-\alpha) \,.
\end{align}

The Fourier transforms of the shape deformations are therefore
\begin{align}
\int \frac{d\theta_1}{2\pi} \lceil \rho_{T_1}\rceil (\theta_1) e^{-i k_1 \theta_1} &= -\int_{-\frac\pi2}^{\frac\pi2} \frac{d\theta_1}{2\pi} \cos\theta_1  e^{-i k_1 \theta_1} 
= \frac{1}{\pi} \frac{1}{k_1^2-1} \cos \frac{\pi k_1}{2} \,, \\
\int \frac{d\theta_2}{2\pi} \lceil \rho_{T_2}\rceil (\theta_2) e^{-i k_2 \theta_2} &= -\int_{-\frac\pi2}^{\frac\pi2} \frac{d\theta_2}{2\pi} \cos\theta_2  e^{-i k_2 (\theta_2+\alpha)} 
= \frac{1}{\pi} \frac{1}{k_2^2-1} e^{-i k_2 \alpha} \cos \frac{\pi k_2}{2}\,. 
\end{align}

Combining these results, the second shape derivative is
\begin{align}
\delta^{(2)}S_\mathrm{reg} 
\Big[ \lceil \rho_{T_1}\rceil , \lceil\rho_{T_2} \rceil \Big]   
&= -\frac{\pi^2}{6} C_T \sum_{k\neq\{-1,0,1\} }  
\frac{|k|}{k^2-1} e^{ik\alpha}   \cos^2 \frac{k \pi}{2}\,,\\
&= \frac{\pi^2}{6} C_T \left[ 1- (T_1 \cdot T_2) \mathrm{arctanh} (T_1 \cdot T_2) \right]\,.
\label{eq:final}
\end{align}
Since this expression is constructed from $SO(d+1,1)$ invariants, this expression is invariant under conformal transformations.

Now consider any ``nice'' family of balls (as defined in \eqref{eqn:norm_R_condition}) and the associated bulk non-extremal surface $N$ constructed as the envelope of the minimal surfaces anchored on the balls.  Above we gave a prescription for calculating  the differential entropy of the family of balls.   Calculationally, this prescription involved picking vectors  $T_{1,2}(\sigma)$ in embedding space that are orthogonal to each other as well as to the ball $B(\sigma)$.  Parametrized in this way, (\ref{eq:final}) is constant for all $\sigma$, i.e. for all balls.  This makes the integral for differential entropy easy to do in the coordinates we have chosen since the only $\sigma$ dependence comes from the measure.  Thus, plugging  into our formula for differential entropy \eqref{eqn:area_diff_ent} gives
\begin{align}
 S_\mathrm{diff} \big[ B(\sigma) \big] = \frac{\pi^3}{12} C_T \int d \sigma \mu(\sigma) \, .
\end{align}
In Section \ref{sec:int_geo} we explained that the integral of $\mu(\sigma)$ necessarily computes the area of the envelope of extremal surfaces $N$.  (This is demonstrated in Appendix~\ref{app:croft} via direct computation.)  In other words,
\begin{align}
 S_\mathrm{diff}\big[ B(\sigma) \big] 
=  \frac{1}{4 G_N} Area(N) 
\end{align}
Thus our conjecture is satisfied for the vacuum state of any three dimensional field theory with a holographic $AdS$ dual.  In excited states the shape derivatives that we need to compute will not generally be  so simple.   Thus, excited states will constitute a critical test of our conjecture.

\section{Discussion}
\label{sec:discuss}
In order to identify an information theoretic quantity dual to the area of non-extremal bulk surfaces in higher dimensions, we proposed a generalisation of differential entropy. We used the fact that the area of a closed surface in the bulk $AdS$ is a finite and coordinate invariant quantity to constrain our proposal and then we checked our proposal for arbitrary surfaces in $AdS_4$. 

\paragraph{Higher dimensions}
In \eqref{eqn:high_dim_diff_ent_discretized} we proposed a discretized formula for differential entropy in higher dimensions.  Taking the continuum limit gives an expression that
involves $d$ shape derivatives. Then following the prescription described above to choose coordinates adapted to the $SO(d+1,1)$ conformal symmetry, our proposal for differential entropy becomes
\begin{align}
S_\mathrm{diff} \big[ B(\sigma) \big] &= \int d\sigma \, \mu(\sigma) \, \delta^{(d)} S_\mathrm{reg} \big[ \lceil \rho_{T_1} \rceil, \ldots, \lceil \rho_{T_d} \rceil \big]\,, \\
\delta^{(d)} S_\mathrm{reg} \big[ \rho_1 , \ldots, \rho_d] &= \int d\theta_1 \ldots d\theta_d \frac{\partial^d S_\mathrm{reg}}{\partial \delta r(\theta_1) \ldots \partial \delta r (\theta_d)} \rho_1(\theta_1) \ldots \rho_d (\theta_d) \,.
\end{align}
Here   $\mu$ is the unique $SO(d+1,1)$ invariant measure on a d-dimensional family of balls $B(\sigma)$ in the field theory, and the $T_i$ are an orthonomal frame for the tangent space to $B(\sigma)$ in embedding space.  In addition, the $\lceil \rho_V \rceil$ are particular shape deformations associated to a vector $V$ (see \eqref{eq:ceilingdef}), and $\partial S_{reg}/\partial \delta r(\theta_i)$ is a regulated shape derivative of the entanglement entropy of a ball with respect to a deformation at the position $\theta_i$ on the ball.
The higher shape derivatives required to evaluate this quantity have not been computed, and it would be interesting to do so.

\paragraph{Monotonicity}
It was shown in \cite{1805.08891} that there are area laws that ensure that bulk surfaces that are the envelope of a family of extremal surfaces obey certain monotonicity properties given reasonable energy conditions.  The extremal surfaces in question terminate on the boundary of the spacetime on a family of balls $B(\sigma)$. If $B(\sigma)$ is a ``nice'' family  (see \eqref{eqn:norm_R_condition}) and $B'(\sigma)$ is another family that is in the ``future'' \cite{1505.05515} of $B(\sigma)$, i.e.,  every $B'(\sigma)$ is contained in some $B(\sigma)$, then the area of the envelope associated to $B(\sigma)$ is smaller than the area of the envelope of $B'(\sigma)$.\footnote{There are many parallels between concepts introduced in \cite{1805.08891} and those discussed in this work: their coarse-graining families are roughly equivalent to our nice families of balls (although they consider general regions and we restrict to balls) and when they say that a family $\tilde F$ is IR coarser than $F$ we would say that $\tilde F$ is in the future of $F$. We adopt this causal language because it ties in to earlier work describing the causal structure on $\cK$ \cite{1505.05515}.}

The differential entropy was shown to obey a similar monotonicity property for two dimensional CFTs in \cite{1805.08891} directly using strong sub-additivity, although its interpretation in terms of constrained state merging given in \cite{1410.1540} already guaranteed this.  It is not immediately obvious how to extend such a direct proof to our higher dimensional proposal, either as a general property of quantum states or by restricting ourselves to holographic states and using the techniques of re-gluing surfaces introduced in \cite{0704.3719}. However, given the structure of our proposal, it seems plausible that it can be given an interpretation in terms of constrained state merging. 

Specifically, the discretized differential entropy for a two-dimensional field theory can be written in terms of conditional entropies as 
\begin{align}
S_{diff}^{(d=1)} = \sum_i \left[ S(B_i) - S(B_i \cap B_{i-1})  \right]
= \sum_i \left[ S(B_i - B_{i-1} |B_i  \cap  B_{i-1} ) \right] \,,
\end{align}
where $B_i - B_{i-1}$ denotes the difference of sets and $S(A | B) = S(A\cup B) - S(B)$.   Each term in the sum measures the information in a new region $B_i$, conditioned on knowledge of the the part of $B_i$ that overlaps with the region $B_{i-i}$.  This leads to the interpretation that the differential entropy in two dimensions is associated to a constrained state merging protocol in which each ball is added sequentially to construct the global state from the local density matrices.  The differential entropy is the entanglement cost of this state merging protocol -- essentially the number of Bell pairs required to carry it out.

For three-dimensional field theories, the differential entropy that we have proposed can also be written in terms of conditional entropies:
\begin{align}
S_{diff}^{(d=2)} =  \sum_{i,j} \Bigg[ S&(B_{i,j})  
-  S(B_{i,j}\cap B_{i-1,j}) - S(B_{i,j}\cap B_{i,j-1}) \\
&+ S(B_{i,j}\cap B_{i,j-1}\cap B_{i-1,j}\cap B_{i-1,j-1}) \Bigg]\,, \\
= \sum_{i,j} \Bigg[ S&(B_{i,j} - B_{i-1,j} | B_{i,j}\cap B_{i-1,j}) \\
-  S&(B_{i,j}\cap B_{i,j-1} -  B_{i-1,j}\cap B_{i-1,j-1}
| B_{i,j}\cap B_{i,j-1}\cap B_{i-1,j}\cap B_{i-1,j-1}))\Bigg] \,.
\end{align}
This expression resembles the two-dimensional case above, except that there is a double sum.  We can imagine treating this sum ``row-by-row'', i.e. summing over $i$ first and then $j$.   This suggests an interpretation in terms of constrained state merging in which balls are added sequentially row-by-row to construct the global state from the local density matrices.  If such an interpretation is sustained, moving the family of balls towards the ``future'' in our terminology would correspond to imposing a stronger constraint on the state merging task. This would ensure that the entanglement cost  of the merging increases; the differential entropy measures this cost.   An interpretation of differential entropy as bulk area would therefore imply that the area of the bulk surface corresponding to a family of larger boundary balls would necessarily be smaller.  

\paragraph{c-functions in arbitrary dimension}
If our conjecture holds relating differential entropy in arbitrary states to the area of dual surfaces, then \cite{1805.08891} implies new monotonicity properties of the shape derivatives of entanglement entropy. In particular, we would expect the shape derivatives appearing in the differential entropy to be positive and monotonic under rescaling of the size of the balls used to define them. This provides a new proposal for a c-function in arbitrary dimensions.   To establish such a c-function we need to  understand the higher shape derivatives we have considered in this paper to see if they can extract universal terms in the entanglement entropy.

\paragraph{Entropy vs. differential entropy}
We have proposed a formula for differential entropy as the dual description of the area of non-extremal surfaces in higher dimensional $AdS$ space.  The areas of {\it extremal} surfaces like the RT-surfaces and the horizons of black holes are holographically dual to von Neumann entropies.  The differential entropy is not known to have such an interpretation; rather it measures the ignorance of local observers of a global state as quantified by a constrained state merging protocol.  We could ask what our formula would give if we evaluated it for an extremal surface.  In that case, our expressions reduce to boundary terms which we have not analyzed in this paper.  These boundary terms evaluated for extremal surfaces should compute von Neumann entropies as they do in the previously understood case of  two dimensional field theories.

\paragraph{$1/N$ corrections}
It is tempting to think of the differential entropy as an actual entropy associated to a time strip in the boundary field theory.  To make sense of this idea we must clearly define entropies that make sense for time strips of a field theory.   An algebraic definition is challenging because  the operators in a time strip do not form a closed algebra. Some work has appeared discussing notions of entropies for subsets of operators which are not closed algebras, but it is not yet clear what is the correct definition \cite{1712.09365,1806.02871}.   It may be better to privilege the interpretation of differential entropy in terms of a state merging protocol. This interpretation tells us that the area of a non-extremal bulk surface computes the cost in Bell pairs of reconstructing the global state of a system from the reduced density matrices of observers restricted to the time strip. This interpretation seems like it could be made well defined in a field theory and could be thought of as quantifying the information which ultraviolet observers confined to the time strip cannot access. In the bulk, this quantifies the information contained in the region not probed by these observer's entanglement wedges.  One could imagine testing this idea by considering the addition of $O(1)$ perturbations in the inaccessible regions.  These should change the differential entropy by a subleading amount that corresponds to the  ignorance of the information contained in the added perturbations in analogy with \cite{1307.2892}.   Computing these $1/N$ corrections will be useful for understanding what differential entropy should mean in quantum gravity.

%

\section*{Acknowledgements }
We thank Jake Bian for early collaborations on related material.  We also thank Bartek Czech, Matt DeCross, Michal Heller, Arjun Kar, Nima Lashkari, Aitor Lewkowycz, Rob Myers, Onkar Parrikar, Mark Van Raamsdonk, Phillippe Sabella-Garnier and G\'{a}bor S\'{a}rosi for discussions.   VB and CR acknowledge support from the Simons Foundation (\# 385592, VB) through the It From Qubit Simons Collaboration, and the US Department of Energy contract \# FG02-05ER-41367.  CR additionally acknowledges support from the Belgian Federal Science Policy Office through the Interuniversity Attraction Pole P7/37, by FWO-Vlaanderen through projects G020714N and G044016N, and from Vrije Universiteit Brussel through the Strategic Research Program ``High-Energy Physics''.  We thank the Galileo Galilei Institute for Theoretical Physics, Florence for their hospitality during the workshop ``Entanglement in quantum systems'', where part of this work was carried out.    CR thanks the Perimeter Institute for Theoretical Physics\footnote{Research at Perimeter Institute is supported by the Government of Canada through the Department of Innovation, Science and Economic Development and by the Province of Ontario through the Ministry of Research and Innovation.} as well as the Mainz Institute for Theoretical Physics during the workshop ``Modern Techniques for CFT and AdS" for hospitality.  VB thanks the Aspen Center for Physics which is supported by National Science Foundation grant PHY-1607611.

\appendix

\section{Double-fibrations for Lorentzian manifolds}
\label{sec:lorentzian}
In this appendix, we will give a covariant version of the  the construction in Section~\ref{sec:maths}. 

Consider an asymptotically-$AdS_{D+1}$ manifold, $M$. A spacelike co-dimension 2 extremal surface in $M$ (a space of signature $(D,1)$) has a normal space of signature $(1,1)$. Instead of a normal vector at each point on the surface, there is a normal 2-plane. This normal 2-plane can be parametrized by two null vectors, $N_+$ and $N_-$ chosen such that $g_{\mu\nu} N_+^\mu N_-^\nu=1$ modulo the rescaling 
\begin{align}
 N_+ \rightarrow \lambda N_+ \qquad N_- \rightarrow \lambda^{-1} N_- \,.
\end{align}
 It may also be specified by an orthogonal spacelike unit normal $n_s$ and timelike unit normal $n_t$ modulo a boost in the normal plane. The specification of such a 2-plane locally at a point by either of these approaches will be referred to as a 2-frame. We will denote the bundle of 2-frames on $M$, the 2-frame bundle, as $\mathbb{F}_2 M$. This is analogous to the bundle of unit vectors we defined in the text, which could be thought of as the 1-frame bundle:
\begin{align}
\mathbb{S}M = \mathbb{F}_1 M \,.
\end{align}

Again, there is a volume form
\begin{align}
vol = \sqrt{\det g} dx_1 \wedge \ldots \wedge dx_{D+1}
\end{align}
on $M$, but the area of a co-dimensions 2 surface $N$ is not given by integrating a form over the surface -- there is an implicit dependence on the 2-frame that can be written as
\begin{align}\label{eqn:Lorentzian_area}
Area(N) = \int_{\tilde N} area \,,\\
area = \iota_{N_-} \iota_{N_+} vol \,,
\end{align}
where $\tilde N$ is again the lift of $N$ into a section of $\mathbb{F}_2 M$ by appending the normal 2-frame at every point of the surface, and $area$ is a form defined on $\mathbb{F}_2 M$.

A covariant generalization of the kinematic space defined in the main text the space of all balls and boosts of balls in the boundary. We will continue to call this space $\cK$.  $\cK$ is closed under the global conformal group and was discussed in detail in \cite{1606.03307}.
Now consider defining $E$ as the bundle over $\cK$ where the fibres consist of the points on the bulk extremal surface anchored to a (boosted) ball shaped region just as before. This bundle $E$ has a natural embedding into $\mathbb{F}_2 M$ by taking the 2-frame normal to the surface at the given point. 

Let us go over this for the example of $M = \mathrm{AdS}_{D+1}$ using Poincar\'e coordinates:
\begin{align}
ds^2 = \frac{-dt^2 + dz^2 + dx_i^2}{z^2}\,.
\end{align}
A co-dimension 2 surface $N$ can be specified locally by two conditions $t=t(x)$ and $z=z(x)$. The 2-frame normal to this surface is then specified by the unit vectors
\begin{align}
n_s= \frac{z}{\sqrt{1+(\partial_{x_i} z)^2}} (0,1,\partial_{x_i} z) \,, 
\qquad n_t=\frac{z}{\sqrt{1-(\partial_{x_i} t)^2}}(1,0,\partial_{x_i} t) \,.
\end{align}
It is therefore natural to consider coordinates on $\mathbb{F}_2 M$ similar to those used for $\mathbb{S}M$  in the body of the text: $(t,z,x_i;\dot t_i, \dot z_i)$  such that the section $\tilde N$ is defined by $\dot z_i = \partial_{x_i} z$ and $\dot t_i = \partial_{x_i} t$.

Now introduce embedding space: for this Lorentzian case the conformal symmetry is $SO(D,2)$, so $AdS_{D+1}$ is the space of vectors in $\mathbb{R}^{D,2}$ such that $X^2=-1$. A 2-frame is specified by $S^2=1$ and $T^2=-1$ such that $S\cdot T=0$ modulo the $SO(1,1)$ boosts in the $S$-$T$ plane or equivalently by $N_\pm = S \pm T$, where the boosts become $N_\pm \rightarrow \lambda^{\pm 1} N^\pm$. As discussed in \cite{1606.03307}, the space of (boosted) balls $\cK$ is parametrized by just such a 2-frame: two vectors $S^2=1$ and $T^2=-1$ up to the $SO(1,1)$ boosts. The vector $T$ can be thought of as specifying the Lorentz frame where the ball lies at constant time and then the vector $S$ in this frame is simply the vector $B$ from before. A point on a ball $(S,T)$ is a vector $X^2=-1$ such that $X\cdot S = X\cdot T=0$. Bringing this together, a point in $E$ is a triple of orthogonal unit vectors $(S,T,X)$ where $S$ is spacelike and $T$ and $X$ are timelike modulo boosts in the $S$-$T$ plane (but not in the $X$-$T$ or $X$-$S$ planes---these correspond to rotating the extremal surface around the given point). A point in $\mathbb{F}_2 M$ is also given by a triple of mutually orthogonal\footnote{$S$ and $T$ must be orthogonal to $X$ because although they are normal to the submanifold $N \subset M$ they are still in the tangent space to $M$.} vectors $(X,S,T)$ modulo boosts in the $S$-$T$ plane. We therefore see that $\mathbb{F}_2 M \cong E$ for the case of empty AdS.

What about the ``null vector alignment condition'' that appeared in the proof of the the relation between two-dimensional differential entropy and the area of bulk curves in \cite{1408.4770}? There it was found that for $D=2$, where the extremal surfaces are curves, the envelope of a family of extremal surfaces may not actually be tangent to a particular extremal surface at each point. Instead, the tangent vector to the curve maybe be shifted by a null vector in the normal 2-plane. Starting from \eqref{eqn:Lorentzian_area}, we will now see that such a shift does not affect the area element of the surface at that point. Consider a surface parametrized by $X^\mu(\sigma_i)$, $i=1,\ldots, D-1$: in terms of the volume form 
\begin{align}
vol = \sqrt{\det g}\,\epsilon_{\mu_1 \ldots \mu_{D+1}} dX^1 \wedge \dots \wedge dX^{D+1}\,,
\end{align}
the area element on the surface is
\begin{align}
\sqrt{\det g}\, \epsilon_{\mu_1 \ldots \mu_{D+1}} N_+^{\mu_D} N_-^{\mu_{D+1}} \partial_{\sigma_1} X^{\mu_1} \ldots  \partial_{\sigma_{D-1}} X^{\mu_{D-1}} d\sigma_1 \ldots d\sigma_{D-1}\,.
\end{align}
Now shift one of the tangents by one of $N_\pm$.  Without loss of generality consider shifting the first one by $N_+$: 
\begin{align}
\partial_{\sigma_1} X^{\mu} \rightarrow \partial_{\sigma_1} X^{\mu} + c N_+^\mu\,\\
N_-^\mu \rightarrow N_-^\mu  - c \partial_{\sigma_1} X^{\mu} +c^2 N_+^\mu \,,
\end{align}
where the second line is the resulting shift in the normal 2-frame. To simplify notation, we have assumed without loss of generality that the parametrization $X^\mu(\sigma_i)$ is such that the tangents $\partial_{\sigma_i} X$ are already orthonormal. By plugging into the above, we see that this does not affect the area element due to the antisymmetry of the $\epsilon$ tensor. A more detailed study will be necessary to understand whether this freedom is sufficient to relate the normal 2-frames of the extremal surfaces in a family to the 2-frames of their envelope.

\section{Crofton-like formula for $AdS_4$}
\label{app:croft}
The logic behind the Crofton formula for computing the length of curves in $AdS_3$  (developed in \cite{1505.05515}) is that in sufficiently symmetric spaces, measures on geometric objects that respect the symmetries are often unique. Some introductory literature on the field of integral geometry, which concerns itself with defining and identifying these measures, can be found at \cite{helgason,santalo_kac_2004,solanes}.


Consider the set of all geodesics which intersect a bulk curve. There is a unique measure on the space of geodesics that is compatible with the isometries of AdS, which restricts to a unique measure on the set of all geodesics intersecting this curve.
From the perspective of the curve, there are geodesics leaving it at every point along the curve in every direction in the unit sphere. Thus an integral over all geodesics intersecting the curve can be split into an integral over the points along the curve times an integral over the unit sphere. Since $AdS$ is isotropic, the integral will be uniform over the unit sphere and its volume will factor out. We have re-expressed an integral over geodesics intersecting a curve in terms of an integral over the points on that curve, but what measure are we integrating over this curve? We know that it must be compatible with the isometries of AdS, and so we must conclude that it is the unique measure on the points along a curve compatible with these isometries: the length of that curve. We must therefore conclude that integrating the unique measure allowed by the isometries of $AdS$ over the space of all geodesics intersecting a curve must give the length of that curve times an overall factor of the volume of a unit sphere.

The same logic applies to integrating over all co-dimension 1 extremal surfaces anchored to boundary balls tangent to a co-dimension 1 surface in a spatial slice of $AdS_4$. This is clearest in embedding space. As reviewed in Section \ref{sec:embed}, the points in $AdS$ are parametrized by timelike unit vectors $X^2=-1$. The unique metric invariant under the isometries of $AdS$\footnote{The language of embedding space makes it easy to construct quantities invariant under the isometries of AdS; we must simply make sure to contract all the $SO(d+1,1)$ indices.} is $ds^2 = dX^2$. This leads to the usual metric on $AdS$ and the usual  measure of area for co-dimension 1 surfaces. The space of balls in a spatial slice of a $CFT_3$ is parametrized by the unit spacelike vectors in embedding space $B^2=1$. Similarly, this space has a unique metric invariant under the conformal symmetry: $ds^2 = dB^2$. This metric is $dS_3$ and when written in coordinates describing the center and radius of the ball it is
\begin{align}
 \frac{-dR^2 + d\vec{x}_0^2}{R^2}\,.
 \end{align}
This leads us to a unique conformally invariant measure on extremal surfaces anchored to balls tangent to a bulk co-dim 1 surface lying on a spatial slice of $AdS_{d+1}$,
\begin{align}
\mu(\vec{x}_0) d^{d-1}\vec{x}_0= \frac{ \sqrt{1- \left( \partial_{\vec{x}_0} R \right)^2 } }{R^{d-1}} d^{d-1}\vec{x}_0,
\end{align}
where we have parametrized the balls by specifying $R(\vec{x}_0)$. By the uniqueness of these measures we expect this measure to be equal to the area of the bulk envelope in $AdS$ -- possibly up to non-unique boundary terms. We will now show this explicitly.

For a spatial slice of $AdS_4$, the measure is 
\begin{align}
\int &\mu(\vec{x}_0) d^{2}\vec{x}_0 \\
= \int &dx_0 dy_0 ~R^{-2} \sqrt{1- \left( \frac{\partial R}{\partial x_0} \right)^2 - \left( \frac{\partial R}{\partial y_0} \right)^2 }. \label{eqn:AdS_result}
\end{align}
where we have denoted $x_0^{(1)}$ by $x_0$ and $x_0^{(2)}$ by $y_0$.

We will now show that this reproduces the area of the corresponding envelope surface in a spatial slice of $AdS_4$. 
The change of variables
\begin{align}
\vec{x} = \vec{x}_0 - R \partial_{\vec{x}_0} R \\
z = R \sqrt{1-(\partial_{\vec{x}_0} R)^2}
\end{align}
 which follows from the equation for a minimal surface anchored on a ball in $AdS$
 \begin{align}
z^2 + \left( \vec{x} - \vec{x}_0 \right)^2 = R^2
  \end{align}
 can be used to write \eqref{eqn:AdS_result} in terms of $z(x,y)$ parametrizing the co-dimension one bulk envelope
\begin{align}
\int dx dy ~
 z^{-2} &\left[ 1+(\partial_x z)^2 + (\partial_y z)^2 \right]^{-\frac32} \\
 \Bigg[&
 1+ (\partial_x z)^2 + (\partial_y z)^2  + z^2 (\partial^2_x z) (\partial^2_y z) - z^2 (\partial_x\partial_y z)^2 \\
 &+ z \partial_x^2 z + z \partial_y^2 z 
 + z (\partial_x z)^2 \partial_x^2 z - 2 z (\partial_x z) (\partial_y z) \partial_x \partial_y z + z  (\partial_y z)^2 \partial_y^2 z
 \Bigg] ,
\end{align}
where we have included the Jacobian for the change of variables $(x_0,y_0) \rightarrow (x,y)$.

This expression is equal to
\begin{align}
\int dx dy ~ \Bigg\{ & \frac{\sqrt{1+(\partial_x z)^2 + (\partial_y z)^2 } } { z^2 } \\
&+ \partial_x \left[ \frac{\partial_x z}{z \sqrt{1+(\partial_x z)^2 + (\partial_y z)^2}} \right]
+ \partial_y \left[ \frac{\partial_y z}{z \sqrt{1+(\partial_x z)^2 + (\partial_y z)^2}} \right] \\
&+ \partial_x \left[ \frac{(\partial_x z) (\partial_y^2 z)}{[1+ (\partial_y z)^2 ] \sqrt{1+(\partial_x z)^2 + (\partial_y z)^2 }  } \right]
- \partial_y \left[ \frac{(\partial_x z) (\partial_x\partial_y z)}{[1+ (\partial_y z)^2 ] \sqrt{1+(\partial_x z)^2 + (\partial_y z)^2 }  } \right]
\Bigg\},
\end{align}
which is the area of a surface in a spatial slice of $AdS_4$ up to boundary terms.

\section{Integral geometry in embedding space}
\label{app:int_geo}
The principles of integral geometry have many applications that were not used in the body of this work. Integral geometry can be described as the field which finds invariant measures on spaces of geometric objects subject to symmetries. Consider the space of straight lines in flat space. These might be parametrized by giving where the line crosses the y-axis and the angle at which it crosses, $(y_0, \theta)$. Subject to the isometries of the plane, which include shifts of $y_0$ and $\theta$, the unique invariant measure is $dy_0  \, d\theta$. This is known as the Crofton form.

The space of points in the plane also has a unique measure compatible with the isometries  of the plane: the familiar Lebesgue measure $dxdy$. The space of points along a curve $\gamma$ in the plane, which can be parametrized by $y(x)$, similarly has a unique measure invariant under these symmetries induced from the measure on the plane
\begin{align}
\sqrt{1+\big(y'(x)\big)^2} dx\,.
\end{align}
The space of geodesics and the space of points along a curve can be related to each other by introducing the space of unit vectors on the plane $\mathbb{SR}^2$, parametrized by $(x,y,\theta)$, which again has a unique measure invariant under the isometries of the plane
\begin{align}
 dx \, dy \, d\theta\,.
\end{align}
This space is a double-fibration: it can be thought of as a bundle in two ways, the projection $\pi_M$ which consists of forgetting the vector and keeping the point and the projection $\pi_\cK$ which consists of shooting out a line in the direction of the vector and forgetting which point along the line we started at. By moving through this auxiliary space, we can relate the two bases. This space space of unit vectors is the analogue of the bundles $\mathbb{S}M$ and $E$ used the body of the text.\footnote{For homogeneous spaces, the bundles of unit vectors $\mathbb{S}M$ and of points along extremal surfaces $E$ are the same. However in general these spaces are not the same and so we had to distinguish them in the body of the text.} 

Consider the space of all lines crossing the curve $\gamma$. In terms of the double-fibrations structure this is $\pi_\cK ( \pi_M^{-1}( \gamma ))$, that is first we take all the whole fibre above each point on the curve, then we project onto the other base. Now consider integrating the unique measure on $\mathbb{SR}^2$ over all the lines intersecting the curve $\gamma$ (in an abuse of notation, we will refer to the measure invariant under the isometries of the plane on whatever space we are integrating over at the moment as $\mu$)
\begin{align}
\int_{\pi_M^{-1}(\gamma) } d\mu = (\int fibres ) \left( \int_\gamma d\mu \right) = 2\pi ~ Length(\gamma) \,,
\end{align}
but this can also be expressed as
\begin{align}
\int_{\pi_M^{-1}(\gamma) } d\mu = \int_{\pi_\cK ( \pi_M^{-1}(\gamma))} (\#~of~intersections) \, d\mu  \,,
\end{align}
where $(\#~of~intersections)$ denotes the number of times a given line intersects with the curve $\gamma$ (for a convex curve $\gamma$ this is zero, one or two).
Putting these together we obtain the Crofton formula:
\begin{align}
2\pi ~ Length(\gamma) =  \int_{\pi_\cK ( \pi_M^{-1}(\gamma))} (\#~of~intersections) \, d\mu \,.
\end{align}

This principle can be generalised to $AdS$; the only challenge is that constructing invariants of the $AdS$ isometry group is non-trivial in Poincare coordinates. This is where embedding space is useful, since it becomes trivial to write down all possible  invariants.
Extremal surfaces and geodesics are simply the restrictions of planes and straight lines respectively to the section of embedding space where $AdS$ lives. A $d-k+1$-dimensional totally geodesic submanifold of a spatial slice of $AdS$ ($\mathbb{H}_{d+1}$)  which is also an extremal surface, can be specified as the space of unit timelike vectors orthogonal to a $k$-plane in embedding space. Given $k$ orthonormal vectors defining a plane in embedding space, $S_1, \ldots, S_k$, the surface consists of the points $X^2=-1$, $X\cdot S_i=0$. The only building blocks available to write a $SO(d+1,1)$ invariant metric on this space of planes are $dS_i \cdot dS_j$ for $i,j=1,\ldots, k$. The requirement for invariance under the $SO(k)$ subgroup acting along the plane gives a unique metric up to rescalings by an overall constant:
\begin{align}
ds^2_{(k)} = \sum_{i=1}^k dS_i \cdot dS_i \,,
\end{align}
leading to a $SO(d+1,1)$ invariant measure on the space of $k$-planes in embedding space or equivalently the space of $d-k+1$-dimensional totally geodesic submanifolds of $\mathbb{H}_{d+1}$. Using similar techniques, we can imagine relating many types of geometric quantities in AdS.

\section{Facts about discretizations}
\label{app:discretization}
We used the fact that for fine enough discretizations
\begin{align}
 \partial B_{i,j} \subset \bigcup_{a\in \{\rightarrow,\leftarrow,\uparrow,\downarrow\} } &B_a \,, \label{eqn:ball_facts_1} \\
 \Big(B_\rightarrow \cap \partial B_{i,j} \Big) \cap \Big( B_\uparrow \cap \partial B_{i,j} \Big) &\neq \emptyset \,, \label{eqn:ball_facts_2}\\
 B_\rightarrow \cap B_\uparrow \cap \partial B_{i,j} &\subset B_{\rightarrow + \uparrow} \,. \label{eqn:ball_facts_3}
\end{align}
In the limit of a fine discretization, these can be reformulated in terms of the shape deformations introduced in Section \ref{sec:discretizations}.   In the limit of a fine discretization ($a \rightarrow 0$), the intersection $B_{\rightarrow} \cap	\partial B_{i,j}$  is the region of $\partial B_{i,j}$ where $\rho_\rightarrow$ is negative (this was discussed above \eqref{diag:inside}). $T_\rightarrow$ and $T_\uparrow$ are orthogonal to each other and to $B_{i,j}$; so they can be completed to an orthonormal basis by adding a unit timelike vector $W$. The space of null rays orthogonal to $B_{i,j}$ can be parametrized by 
\begin{align}
Z(\theta) = W + \cos\theta\, T_\rightarrow + \sin\theta\, T_\uparrow\,.
\end{align}
The shape deformation towards a neighbouring ball is
\begin{align}
\rho_{\{\rightarrow,\leftarrow,\uparrow,\downarrow\}} &= - Z(\theta) \cdot T_{\{\rightarrow,\leftarrow,\uparrow,\downarrow\}} = - \cos(\theta-\theta_0) \,, \\
\theta_0^{\{\rightarrow,\leftarrow,\uparrow,\downarrow\}} &= \{0, \pi, \frac\pi2,\frac{3 \pi}{2} \}\,.
\end{align}
From these shape deformations we can determine that in the limit of a fine discretization ($a \rightarrow 0$),
\begin{align}
B_{\rightarrow} \cap	\partial B_{i,j} &= [ -\frac\pi2,\frac\pi2 ]\,, &\qquad B_{\leftarrow} \cap	\partial B_{i,j} &= [\frac\pi2,\frac{3\pi}2 ] \,,  \\
B_{\uparrow} \cap \partial B_{i,j} &= [ 0, \pi ]\,, &\qquad B_{\downarrow} \cap	\partial B_{i,j} &= [\pi,2\pi ] \,.
\end{align}
As we back off from the strict $a\rightarrow 0$ limit, these regions will continuously shrink at their boundaries.  From these explicit expressions, equations  \eqref{eqn:ball_facts_1} and \eqref{eqn:ball_facts_2} are clear.

In the limit of a fine discretization, $B_{\rightarrow + \uparrow} \cap \partial B_{i,j}$ is the region where the shape derivative is negative (again see discussion above \eqref{diag:inside}),
\begin{align}
\rho_{\rightarrow + \uparrow} &= - (T_\rightarrow + T_\uparrow) \cdot Z = - \sqrt{2} \cos \frac{\pi}{4} \,, \\
B_{\rightarrow + \uparrow} \cap \partial B_{i,j} &= [ -\frac{\pi}{4}, \frac{5\pi}{4} ] \,.
\end{align}
Since the fine discretization limit of $B_\rightarrow \cap B_\uparrow \cap \partial B_{i,j}$ is contained in the interior of this region, \eqref{eqn:ball_facts_3} will hold for sufficiently fine discretizations.

%

\providecommand{\href}[2]{#2}\begingroup\raggedright\endgroup

\end{document}